\documentclass[a4paper,11pt]{article}
\pdfoutput=1
\usepackage{jcappub}
\usepackage{afterpage,booktabs}
\usepackage{graphicx,subfig}
\usepackage{amssymb,amsmath,bm}
\usepackage{mathtools}
\usepackage[usenames,dvipsnames]{xcolor}
\usepackage[capitalise]{cleveref}
\usepackage{expl3,xstring}
\usepackage{dcolumn}
\usepackage{bm}
\usepackage{color}
\usepackage{breqn}
\usepackage{multirow}
\usepackage[normalem]{ulem}
\usepackage{tabu}
 \usepackage{pifont}
\usepackage{xcolor}
\usepackage{fontawesome}
\usepackage{comment}
\usepackage{float}

\definecolor{earlyde}{RGB}{255, 192, 203}
\definecolor{unphysical}{RGB}{106,90,205}
\definecolor{nomatterdom}{RGB}{218,112,214}
\definecolor{viable}{RGB}{0,0,128}
\definecolor{star}{RGB}{255,165,0}
\definecolor{royalgreen}{RGB}{19,120,7}

\newcommand{\HiCOLA}{\texttt{Hi-COLA}}

\title{\boldmath Cubic Galileon Gravity on Nonlinear Scales with \texttt{Hi-COLA} for LSST Year 1}

\author{C. M. A. Zanoletti$^1$,}
\author{A. Sen Gupta$^{2,3}$,}
\author{N. Ramachandra$^{4}$,}
\author{T. Baker$^3$,}
\author{B. Fiorini$^3$,}
\author{K. Heitmann$^6$,}
\author{D. Jones$^{8,9}$,}
\author{C. D. Leonard$^1$,}
\author{B. Li$^5$,}
\author{N. \v{S}ar\v{c}evi\'{c}$^7$,}
\author{for the LSST Dark Energy Science Collaboration}

\affiliation[1]{School of Mathematics, Statistics and Physics, Newcastle University, Newcastle upon Tyne, United Kingdom}
\affiliation[2]{Faculty of Physics, Bielefeld University, Postfach 100131, 33501 Bielefeld, Germany}
\affiliation[3]{Institute of Cosmology and Gravitation, University of Portsmouth, \\ 
Burnaby Road, Portsmouth PO1 3FX, UK}
\affiliation[4]{Computational Science Division, Argonne National Laboratory, 9700 S Cass Ave, Lemont, IL 60439, USA}
\affiliation[5]{Institute for Computational Cosmology, Department of Physics, Durham University, South Road, Durham DH1 3LE, UK}
\affiliation[6]{High Energy Physics Division, Argonne National Laboratory, 9700 S Cass Ave, Lemont, IL
60439, USA}
\affiliation[7]{Department of Physics, Duke University, Science Dr, Durham, NC 27710, USA}
\affiliation[8]{Department of Physics, University of Michigan, Ann Arbor, MI, USA}
\affiliation[9]{Leinweber Institute for Theoretical Physics, Ann Arbor, MI, USA}

\emailAdd{c.m.a.zanoletti2@newcastle.ac.uk}\emailAdd{asengupta@physik.uni-bielefeld.de}
\emailAdd{bartolomeo.fiorini@port.ac.uk}
\emailAdd{tessa.baker@port.ac.uk}
\emailAdd{dericj@umich.edu}

\date{\today}

\abstract{
Vera C. Rubin Observatory's Legacy Survey of Space and Time (LSST) is in a prime position to test Modified Gravity (MG) with Weak Lensing (WL) and Large Scale Structure (LSS). However, many nonlinear MG emulators in the literature enforce a $\Lambda$CDM background by design, missing where data suggest the most interesting deviations lie. We present an emulator for the 3D nonlinear matter power spectrum in Cubic Galileon gravity, a Horndeski theory with a single additional MG parameter,  $f_\phi$, that modifies the background, growth, and lensing. Trained on \texttt{Hi-COLA} simulations with Gaussian Process Regression (GPR), the emulator achieves $\lesssim 2\%$ accuracy for $k \leq 1.2\ h\,\text{Mpc}^{-1}$ at the fiducial cosmology relative to \texttt{Hi-COLA}, and $\lesssim 5\%$ relative to full N-body \texttt{EFT-RAMSES} simulations. Additionally, conservative scale cuts excluding angular bins affected by baryonic physics enable unbiased parameter estimation in LSST Year 1 weak lensing and galaxy clustering ($3\times2$pt) forecasts, even when validated against simulated data based on higher-resolution N-body simulations. This demonstrates that, within a simplified LSST Y1 setup and the adopted scale cuts, \texttt{Hi-COLA}-based emulation is sufficient to develop inference infrastructure with Stage IV precision for Horndeski gravity models previously unmodeled on nonlinear scales.}

\begin{document}
\maketitle

\section{Introduction}

In the standard cosmological model, $\Lambda$CDM, the dark energy responsible for the Universe's accelerated expansion is described by a cosmological constant -- an assumption that can explain a suite of cosmological observations, including Type Ia supernovae (SN Ia) distances \cite{Scolnic_2018}, anisotropies of the cosmic microwave background \cite{2020}, and data from a range of Weak Lensing and Large Scale Structure surveys \cite{PhysRevD.103.083533}. Nonetheless, the cosmological constant comes with significant theoretical challenges, and recent Baryon Acoustic Oscillation (BAO) results from the Dark Energy Spectroscopic Instrument (DESI) \cite{garciaquintero2025cosmologicalimplicationsdesidr2} and multi-probe constraints from the full six-year Dark Energy Survey (DES Y6) \cite{descollaboration2026constraintsdynamicaldarkenergy} are in tension with the standard cosmological model's picture of dark energy. We need to test extensions to $\Lambda$CDM to assess whether these tensions can be explained by new physics, or whether we must instead keep looking for misspecified modelling in the systematics sector of the model. 

Driven by theoretical and observational limitations of GR, a well-motivated class of extensions can be found in modified gravity theories \citep{clifton2012modified, Koyama_2016,Ishak_2018}. It is possible that the observed acceleration is in fact a signature of a departure from General Relativity (GR) rather than the presence of an additional dark energy fluid. In some modified gravity models, an effective dark energy equation of state $w(a)$ that evolves with scale factor $a$ can arise naturally, and can even cross the so-called phantom divide, the boundary at $w=-1$ between the quintessence ($w>-1$) and phantom ($w<-1$) regimes \citep{chudaykin2024modifiedgravityinterpretationevolving, yao2025generalmodeldarkenergy, Wolf_2025a, Wolf_2025b, ye2026latetimereconstructionnonminimallycoupled, Wolf_2026, linder2025cosmologyphantomcrossinghorndeski, calderon2026chargingphantomdividemodified, Tsujikawa_2026, Ye_2025}. In this sense, modified gravity provides a physically richer and more predictive explanation for time-evolving cosmic acceleration than a DE fluid. For example, many models belonging to the Horndeski class alter the cosmological background as well as the growth of structure and lensing of light \citep{Bellini_2014}, including with the \HiCOLA{} code used in this work \cite{Naidoo:2026umv}. 
This scenario exposes a gap in the existing MG modelling landscape: Hu-Sawicki $f(R)$ \cite{PhysRevD.76.064004} and nDGP gravity \cite{DVALI2000208}, the two theories of gravity for which nonlinear simulation and emulation infrastructure is most developed (see e.g., \href{https://bitbucket.org/arnoldcn/forge\_emulator/}{\texttt{FORGE-Emu}}, \href{https://github.com/renmau/Sesame\_pipeline}{\texttt{Sesame}}, \href{https://github.com/HAWinther/FofrFittingFunction}{\texttt{FofrFittingFunction}}, \href{https://pypi.org/project/emantis/}{\texttt{e-MANTIS}} and \texttt{BRIDGE} \cite{harnoisdéraps2022mglensmodifiedgravityweak}), are conventionally implemented with a cosmological background fixed to that of $\Lambda$CDM, precisely where much of the currently interesting deviation from the standard cosmological model lies. There is therefore a clear need to develop nonlinear modelling tools for theories that modify the background expansion as well as the growth of structure.

Weak lensing (WL) and Large Scale Structure (LSS) are ideally suited to test deviations to our standard cosmological model, especially beyond modifications to the background evolution. They are exceptionally rich probes because they encode information across a range of redshifts, acting complementarily to probes of the cosmic microwave background (CMB) anisotropies. The galaxy density field and the weak lensing shear field jointly constrain the lensing of light, the clustering of matter, and the background evolution as a function of time. For example, it is standard to use combined constraints from the two-point correlation functions generated by these fields: galaxy clustering, galaxy-galaxy lensing and cosmic shear (jointly called the $3\times2$pt datavector) \cite{descollaboration2026darkenergysurveyyear, Heymans_2021,miyatake2023hypersuprimecamyear3}, though beyond-2pt statistics of these fields exist and have been used for cosmological inference \cite{Charnock_2018, Gough_2022,10.1093/mnras/stab1623,10.1093/mnras/stx2793,Grewal_2022,Valogiannis_2022,Hahn_2021}. In the near future, the first cosmological data from the Vera C. Rubin Observatory's Legacy Survey of Space and Time (LSST) \cite{Ivezi__2019} and Euclid \cite{2025} will deliver competitive constraints on gravity \citep{ishak2019modifiedgravitydarkenergy,Leonard_2026, CASAS201773, Frusciante_2024,euclidcollaboration2025euclidpreparationconstrainingparameterised}.

In order to model accurately the two-point statistics we can extract from these surveys, we need to compute the 3D matter power spectrum on nonlinear scales. This can be done far into the nonlinear regime by using N-body simulations. For the purpose of repeated model evaluation during parameter inference, these unfortunately remain prohibitively computationally expensive; therefore emulators (fast, approximate surrogate modelling tools, usually based on machine learning methods) have been proposed, developed and validated in the literature as a means of quickly replicating the results of these simulations in a fraction of a second \cite{gordon2024modeling}. Emulators are an important example of the novel modelling tools required to incorporate small-scale information into the exploration both of the standard cosmological model (see \href{https://github.com/miknab/EuclidEmulator2}{\texttt{EuclidEmulator2}} and \href{https://bacco.dipc.org/}{\texttt{BACCOemu}}) and of beyond-$\Lambda$CDM models. 

While summary statistics in Hu-Sawicki $f(R)$ and nDGP gravity have been extensively emulated, very few other MG theories have N-body simulation infrastructure readily available, which hinders the production of training data sets on which emulators are built. New tools (which include approximate N-body simulations and theoretical prescriptions) like \HiCOLA\footnote{\url{https://github.com/Hi-COLACode/Hi-COLA}} \cite{Wright_2023} are making it less computationally expensive to estimate non-linear MG power spectra for a broader parameter space, opening up the possibility of systematically building emulators for a wide range of alternative theories of gravity. A range of beyond-$\Lambda$CDM emulators have been built, often based on approximate nonlinear modelling schemes such as the halo model reaction \cite{10.1093/mnras/stac3783} (see e.g., \texttt{ReACT}-based emulators\footnote{\url{https://github.com/nebblu/ReACT-emus}} and \cite{saadeh2025fieldlevelreactionscreenedmodified}), or COLA (see e.g., \href{https://github.com/LSSTDESC/mgemu}{\texttt{MGemu}} and \cite{srinivasan2026cosmologicalgravityscalesv}). For the latter, \cite{fiorini2023fastproductioncosmologicalemulators} provides a framework for building these emulators and additionally introduces \href{https://github.com/BartolomeoF/nDGPemu}{\texttt{nDGP-emu}}. In this paper, we take a significant step in this direction, introducing and validating the first nonlinear matter power spectrum emulator for Cubic Galileon (CG) gravity, accurate enough to meet LSST Year 1 (`Y1' hereafter) requirements.

Here, we present a worked example of how to build a nonlinear emulator and carry out an LSST Y1 forecast for a theory of gravity that does not enforce a $\Lambda$CDM background by construction. Cubic Galileon gravity \citep{Nicolis_2009, Deffayet_2009a,Deffayet_2009b} presents a compelling test case. It has some clear advantages over $f(R)$ and nDGP gravity: unlike these theories, the specific version of Cubic Galileon gravity we targeted when building this emulator does not incorporate extraneous dark energy components. Although the model includes a cosmological constant, its background evolution can deviate from $\Lambda$CDM, avoiding the fine-tuning of an additional dark energy component at the level of the field equations and resulting in a more theory-driven than a phenomenology-driven framework. 
Still, we must be clear that the Cubic Galileon model we consider fares relatively poorly against current observations. It cannot produce a dark energy equation of state that crosses the phantom divide, it is observationally ruled out for $f_\phi=1$ \cite{Barreira_2014} (where $f_\phi$ is the free parameter of the gravity theory), and it is disfavoured by joint constraints from {\it Planck}, DESI BAO, DES SNe, galaxy-ISW cross-correlations and void force profile measurements \cite{Naidoo:2026umv}.
Nevertheless, this model is a good example of a minimal extension to $\Lambda$CDM that simultaneously modifies the background evolution, the clustering of matter, and the lensing of light.
This establishes a template that can be extended to other Horndeski theories with similar freedom in their background evolution.

This paper is organized as follows. In Section \ref{sec:Theory_and_Sims} we introduce Cubic Galileon gravity and the specific model we consider in this paper (with tracker solution on the background), and then describe the \HiCOLA{} code used to generate simulations that will form the training set for our emulator. In Section \ref{sec:DataGeneration} we outline the generation of our training and validation data and the evaluation of the matter power spectrum boosts. In Section \ref{sec:Emulator_development} we describe the construction of the emulator, outlining both the single-snapshot emulation and the interpolation in redshift. In Section \ref{sec:Inference_methods} we introduce the assumptions and modelling choices for the forecasting infrastructure used in this paper when analysing simulated LSST Y1 $3\times2$pt data. In Section \ref{sec:Emu_results} we present our results: we validate the emulator against \HiCOLA{} simulations and against a higher-resolution \texttt{EFT-RAMSES} N-body simulation \cite{Woodcock:2026} (\texttt{EFT-RAMSES} is a modified gravity simulation code derived from \texttt{ECOSMOG} \cite{Li_2012, Li:2013nua}), and assess its performance as a forecasting tool through a simulated LSST Y1 $3\times2$pt parameter inference analysis. Finally, we conclude in Section \ref{sec:Conclusions}.

\section{Theory and Simulations}\label{sec:Theory_and_Sims}

\subsection{Cubic Galileon and the Tracker Solution}\label{sec:CubicGalileon_theory}

Cubic Galileon gravity (without a cosmological constant) has been widely explored \citep{Traykova_2021,Barreira_2013,Nicolis_2009,Deffayet_2009a,Deffayet_2009b} and strongly constrained observationally \citep{Peirone_2018b,Frusciante_2020}, with full N-body simulations available in the literature \citep{Wyman_2013, Barreira_2013b, Zhang_2020}.
Here, we consider the Cubic Galileon Lagrangian with the following form:
\begin{equation}
    \mathcal{L} = \frac{M_P^2}{2}R +M^2_{\Tilde{K}} k_1 X - M_{\Tilde{G}_{3}}M_{s}g_{31}X \square \phi - M_P^2 \Lambda,
\end{equation}
where $R$ is the Ricci scalar; $M_P$ is the Planck mass; $M_{s}$, $M_{\Tilde{K}}$ and $M_{\Tilde{G}_{3}}$ are the mass scales associated with the scalar field and with the Horndeski functions ${K}$ and ${G}_{3}$; and $X = -\frac{\partial_\mu \phi \partial^\mu \phi}{2}$ is the dimensionless kinetic term of the dimensionless scalar field $\phi$. $k_1$ and $g_{31}$ are dimensionless constants, and we have also included a cosmological constant term $\Lambda$. If we assume all the mass scales are the same (see Appendix A.2 in \cite{Wright_2023}) we can find the following equations of motion at the background level:

\begin{equation}\label{eq:CuGal_background}
    \begin{split}
        \Omega_\phi &= \frac{k_1{\phi'}^2}{6} + g_{31} E^2 {\phi'}^3 \\
        \frac{E'}{E} &= \frac{k_1 {\phi'}^2}{4} - \frac{g_{31}E^4{\phi'}^4}{4}\left(\frac{E'}{E}{\phi'} +\phi''\right) - \frac{\Omega_r -3\Omega_\Lambda + 3}{2} \\
        \phi'' &= -\frac{3g_{31}E'E{\phi'}^2 + 3k_1\phi'+9g_{31}E^2{\phi'}^2}{k_1 + 6g_{31}E^2\phi'} - \frac{E'}{E}{\phi'}
    \end{split}
\end{equation}
where we define the dimensionless Hubble rate $E = H/H_0$, $\Omega_{\phi}$ is the energy contribution of the scalar field (with $\Omega_{\text{DE}} \equiv \Omega_{\Lambda} + \Omega_\phi$), all other quantities have their standard meanings and the prime denotes the derivative with respect to the dimensionless time variable $x$, defined as the logarithm of the scale factor $x = \text{ln}(a)$.

The rescaling symmetry of the Cubic Galileon \cite{De_Felice_2010, De_Felice_2011, Nesseris_2010} gives one the freedom to fix either the initial conditions for $E$ and $\phi'$, or the values of the model constants. For example, one can avoid having to deduce a suitable initial condition\footnote{Whilst we write ``initial'' condition, we mean this term in the sense of a boundary condition, which in this work we specify at late times.} for the scalar field derivative $\phi'$ by rescaling it by its present day value. If we denote quantities that are rescaled under this `today' convention with a $T$ subscript, the initial condition for the Hubble rate and scalar field derivative becomes $E_0 = 1$ and $\phi_{T0}' = 1$ respectively, and accordingly one can redefine the model constants $k_{1T} = k_1(\phi'_{0})^2$ and $g_{31T} = g_{31}(\phi'_{0})^3$ such that the background equations of motion remain invariant. This is the convention we have chosen to adopt for this work (see \cite{Wright_2023} for a more in-depth exploration of the `today' formalism).

Cubic Galileon gravity has a set of stable solutions with de Sitter (dS) attractors, that asymptotically approach a universe dominated by a constant vacuum energy at late times. Regardless of their initial conditions, these solutions converge onto a common trajectory in their background evolution, known as the tracker solution \citep{Felice_2012}. We set up our emulator to produce power spectra for these solutions only. In the `today' formalism, the tracker solution exists on the hypersurface
\begin{equation}\label{eq:tracker}
    k_{1T} = 6 f_{\phi} (\Omega_{m0} + \Omega_{r0} -1)
\end{equation}
of our parameter space, where $f_{\phi}$ is defined as the fraction of the dark energy component due to the scalar field to the total dark energy today ($f_\phi = \frac{\Omega_{\phi 0}}{\Omega_{\phi 0} + \Omega_{\Lambda 0}}$). The initial conditions on $E$ and $\phi_T'$ also mean that the two constants $k_{1T}$ and $g_{31T}$ are dependent, so knowing the value of one is sufficient to find the other (through equations \ref{eq:CuGal_background}). Because we consider only tracker solutions for our emulator, we therefore recover a formalism of the theory described by a single modified gravity parameter $f_\phi$. By varying $f_\phi$, we get a different background history in the recent universe compared to that we expect from $\Lambda$CDM, as well as a different coupling to matter in the Poisson equation and linear growth equation, commonly parameterised by the factor $\mu$ \citep{Jain_2008,Zhao_2009,Pogosian_2016,Baker_2015} (defined in Section \ref{sec:Inference_methods} below). Choosing to restrict to the tracker solution limits generality, but we make it for practical reasons. Off the tracker, $k_{1T}$ and $g_{31T}$ are no longer related by Eq. \eqref{eq:tracker}. The background then depends on a second free parameter, with a correspondingly larger training set, along with requiring stability checks across the extended space. The tracker family is the minimal one-parameter extension of $\Lambda$CDM within this model, and it is therefore the natural first target for the emulation and inference infrastructure developed here. Refs. \cite{Leloup_2019, Neveu_2014, Neveu_2013} confronted the general (non-tracker) solutions of the Galileon model with cosmological data.

By adding a cosmological constant and imposing the tracker solution on the background, we can build an emulator for a viable theory of gravity with a single additional parameter compared to $\Lambda$CDM, whose power spectra can be emulated continuously up to the $\Lambda$CDM limit. Throughout this paper, we make use of two cosmological models, defined as follows. The first is the standard $\Lambda$CDM model built on GR (which includes cold dark matter, radiation including massless neutrinos, and baryons). 
The second is the Cubic Galileon model (based on CG gravity) for which we build our emulator, with the background fixed to the tracker solution described above, and with $\Lambda$CDM initial conditions. The emulated power spectra we use to compute the boost come from \HiCOLA{}, which does not fully include the effects of radiation directly in its forward evolution. We partly compensate for the effect of radiation by using a backscaling approach (described in more detail in \cite{Wright_2023}). For the forecasts in Section \ref{sec:forecasts}, the effects of radiation and baryons are included only insofar as they impact the $\Lambda$CDM power spectrum, which is multiplied by the CG boost (defined in equation \ref{eq:Boost}) to compute the CG power spectrum; we do not expect the boost predictions to be significantly affected by these additional parameters. Massive neutrinos are not included in the present analysis. Because their imprint on the power spectrum cannot in general be decoupled from that of modified gravity (see Fig. 7 of \cite{Mead_2016}), a joint treatment is well-motivated, though on the mildly nonlinear scales probed by our forecasts we expect the impact to be modest. A treatment of massive neutrinos will be implemented in the next version of \HiCOLA{} (in prep.), so future extensions of this emulator to models whose background evolution can cross the phantom divide, including to Asymptotic Cubic Galileons (ACG, \cite{Naidoo:2026umv}) or Rolling Galileons \cite{Hallam:2026qsk}, will be able to include massive neutrinos self-consistently.

\subsection{The \HiCOLA{} code}

Horndeski-in-COLA (\HiCOLA\footnote{\url{https://github.com/Hi-COLACode/Hi-COLA}})~\cite{Wright_2023} is a hybrid particle-mesh (PM) code, which utilises the Co-moving Lagrangian Acceleration (COLA) \cite{COLA_Tassev:2013} scheme to simulate the motion of dark matter particles under gravity more rapidly than traditional $N$-body codes. To give a summary of COLA's key idea: it takes advantage of the known analytic solutions for the large-scale clustering of matter through Lagrangian Perturbation Theory (LPT). 
The PM code solves for the residual displacement between the (second-order) LPT observer and the particles, by performing a boost to the frame of observers that move according to LPT. The resulting numerical solution of the particle geodesic equation converges faster than if the geodesic equation was solved in the initial rest frame of the particles. 
This allows for simulations that can run with fewer time steps while remaining accurate in the mildly nonlinear regime. \HiCOLA{} adapts COLA for use in modified gravity by including the effects of the scalar field on the expansion history and the linear growth factors. In the case of the latter, this leads to a modification of the growth factor equations, as detailed in Section 4.2 of \cite{Wright_2023}. For theories like the Cubic Galileon, which feature scale-independent growth factors, solving these equations remains a computationally inexpensive exercise (see \cite{Valogiannis_2017} for a discussion on dealing with the computational expense of this exercise for theories with scale-dependent growth, such as $f(R)$).

\HiCOLA{} further includes the nonlinear impacts of the scalar field on structure formation through a fifth force (building on \cite{Kimura_vainshtein:2012,Lombriser_MGforce:2016}), making use of the quasi-static approximation (QSA) and weak field approximation \cite{Sirera:2026klo}. 
\HiCOLA{} uses the screening factor approach of \cite{WintherFerreira:2015_FastRouteLSS}. This work showed that the force modification factor derived in a spherically symmetric configuration can be imported to modify the inter-particle forces in a general simulation, with a minimal loss of accuracy at scales $k<1 h\,\text{Mpc}^{-1}$.
This leads to a force expression that can be computed without solving the scalar Klein-Gordon equation, the latter of which is one of the reasons why MG is slower to simulate than GR (see \cite{WintherFerreira:2015_FastRouteLSS} for a discussion). 

Where \HiCOLA{} differs from some of its contemporaries such as \texttt{MG-PICOLA}~\cite{Winther_MGPico:2017}, \texttt{COLA-FML}~\cite{Brando:2022_COLA_emulator}, \texttt{MG-evolution}~\cite{Hassani_MGevo:2020} and the $f(R)$ COLA code of \cite{Valogiannis_2017} is the use of symbolic manipulation to enable a construction of the background equations of motion and approximate fifth force for a generic choice of Horndeski functions ($K, G_3, G_4$). Thus, \HiCOLA{} is designed as a code that is immediately applicable to a wide class of Horndeski theories, rather than specifically addressing select members of the class; and consistently solves for theories from their Lagrangian formulation, instead of relying on approximate parametrisations of the theory space. Currently, \HiCOLA{} has been used to study the clustering phenomenology of Horndeski theories with Vainshtein screening~\cite{Wright_2023} and K-mouflage gravity~\cite{SenGupta:2024_kmou,Bose_code_comparison:2024}, as well as finding viable models with phantom-crossing behaviour \citep{Naidoo:2026umv, Hallam:2026qsk}.

\section{Data Generation for Emulator Development and Validation}\label{sec:DataGeneration}

\subsection{Emulator Training Set: experimental design}\label{sec:Emu_design}
In this work we run simulations for both the $\Lambda$CDM model and the Cubic Galileon model described in Section \ref{sec:CubicGalileon_theory}. We will be emulating the Cubic Galileon boost, defined as:
\begin{equation}\label{eq:Boost}
    B(z,k, \mathbf{p}) = \frac{P_\delta^{\text{CuGal}}(z,k, \mathbf{p})}{P_\delta^{\Lambda\text{CDM}}(z,k, \mathbf{p})}
\end{equation}
where $P_\delta$ is the 3D matter power spectrum computed with \HiCOLA{} and the labels stand for the Cubic Galileon or $\Lambda$CDM model. This quantity will be computed as a function of wavenumber $k$, redshift $z$ and a set of parameters $\bold{p} = \{\Omega_{m0}, n_s, A_s, h, f_\phi\}$, over the ranges:
\begin{equation}
    \begin{split}
        &0.275 \leq \Omega_{m0} = \Omega_{c0} + \Omega_{b0} \leq 0.33 \\ 
        &0.85 \leq n_s \leq 1.1 \\
        & 1.45 \times 10^{-9} \leq A_s \leq 3.3 \times 10^{-9} \\
        &0.61 \leq h \leq 0.73 \\
        &0.02 \leq f_\phi \leq 1
    \end{split}\label{eq:param_ranges}
\end{equation}
where $\Omega_{m0}$ is the ratio of today's total matter density to today's critical density of the Universe and $h$ is the dimensionless Hubble constant defined by $H_0 = 100\,h\, \text{km s}^{-1}\text{Mpc}^{-1}$, with Hubble parameter today, $H_0$, in $\text{km s}^{-1}\text{Mpc}^{-1}$. Additionally, $A_s$ is the primordial super-horizon amplitude of curvature perturbations at $k = 0.05$  $h\,\text{Mpc}^{-1}$, $n_s$ is the primordial spectral index, and $f_{\phi}$ is the additional Cubic Galileon parameter described in Section \ref{sec:CubicGalileon_theory}. We compute the boost with constant $\omega_r = \Omega_r h^2 = 4.18 \times 10^{-5}$, equivalent to assuming a constant $T_{\rm CMB} = 2.725 \text{K}$ and massless neutrinos with $n_{\text{eff}}=3.044$. We also fix the physical baryon density $\omega_b = \Omega_b h^2 = 0.0223$, assuming the boost is insensitive to its variation. This is motivated by Figure 10 in \cite{Heitmann_2009}, which shows that the $\Lambda$CDM power spectrum is largely unaffected by $\omega_b$ (assuming tight priors on this parameter from CMB constraints). We extend this assumption to a beyond-$\Lambda$CDM model, and justify this extrapolation in part because modifications are expected to be smaller for the boost than for the power spectrum (see e.g. Fig. 1 in \cite{S_ez_Casares_2023}). Selection of the ranges of parameters depends on the desired emulator-accuracy requirements and the coverage required for exploration during the Bayesian calibration process. Our cosmological parameter ranges are consistent with those of Coyote Universe \cite{Heitmann_2009} and Mira-Titan Universe simulation suites \cite{Heitmann_2016},  where the precision goals of CMB and large-scale-structure surveys guide the choices. 
The modified-gravity parameter, by contrast, is sampled over the full range permitted by the theory. This extends beyond the limit imposed by current constraints on Cubic Galileon gravity, but it is standard practice to train MG emulators over a wide MG parameter range. For example, Hu-Sawicki $f(R)$ is constrained in Refs. \citep{PhysRevD.102.104060,10.1093/mnras/stae2096} to $\text{log}_{10}|f_{R0}| \,\lesssim -8$ at the 95\% C.L., but emulators for this theory have been trained up to $f_{R0}\sim 10^{-4}$ \cite{Arnold_2022, S_ez_Casares_2023, Ramachandra_2021}.  The reason is that an emulator is used for more than deriving constraints for the specific modified gravity theory, for example serving in pipeline validation and as a building block for alternative modelling schemes \cite{Euclid_2025_beyondLCDM, Zanoletti_2025}.

Our sampling strategy involves finding these parameters at 50 points spaced uniformly in parameter space using a 5D Symmetric Latin Hypercube (SLH) \cite{Heitmann_2009} sample. Alternative Space-Filling Sampling (SFS) methods include low-discrepancy sequences (e.g., Sobol’s sequence). These have the advantage of not requiring a predetermined training sample size; thus, increasing the size of the training set does not require generating an entirely new sample. We opt for an SLH sample because, although it is difficult to determine an appropriate training sample size in advance, a very similar emulation setup was used in \cite{Ramachandra_2021} for comparable purposes, offering a reasonable starting point for subsequent validation in our case. These SLH sample training points for the emulator are displayed in blue in Figure \ref{fig:Validation_param_set}. 

\subsection{Emulator Training Set: evaluating the boosts} 

\begin{table}[]
    \centering
    \begin{tabular}{c|c}
    \hline
         Box Size, $L$ & 400 $\text{Mpc}/h$ \\
         Number of particles, $N_{\text{part}}$ & $512^3$ \\ 
         IC grid size, $N_{\text{mesh}}$ & $512^3$ \\ 
         Grid size for computing particle mesh forces, $N_{F,\text{mesh}}$ & $(3\times512)^3$ \\ 
         Number of time steps, $N_{t}$ & $51$ \\
         Initial Redshift, $z_{i}$ & $49$ \\ 
         Final Redshift, $z_{f}$ & $0$ \\
         \hline
    \end{tabular}
    \caption{Specifications for the \HiCOLA{} training simulation suite.}
    \label{tab:Sim_specs}
\end{table}


We compute the power spectra using the \HiCOLA{} software, following the specifications from Table \ref{tab:Sim_specs}, for both the $\Lambda$CDM and the Cubic Galileon models with the same initial random seed. In order to account to some degree for the effect of radiation in the simulations, the initial conditions were generated from power spectra at $z=0$ that include a radiation component $T = T_{\text{CMB}}$, which were then back-scaled to $z=z_i$ using growth factors under the assumption of zero background radiation. We assume that the GR and Cubic Galileon initial conditions are the same at $z=49$, and use a $\Lambda$CDM matter power spectrum to generate the initial fields for both of these models.

For each node in our training set we run two simulations with initial modes exactly out of phase - that is, with matching amplitudes but phase-reversed initial conditions \cite{R_cz_2023}. We do this in order to efficiently remove finite-volume effects from power spectrum estimates \cite{PhysRevD.93.103519}. We validate this approach by comparing to the average over 10 random realizations and selecting the initial condition seed that (averaged between the two phase-reversed realizations) minimizes the error. The resulting boost averaged over the paired realizations:

\begin{equation}\label{eq:Boost_phasepair}
    B(z,k,\mathbf{p}) = \frac{P_{\delta, \text{Phase 1}}^{\text{CuGal}}(z,k, \mathbf{p}) + P_{\delta, \text{Phase 2}}^{\text{CuGal}}(z,k, \mathbf{p})}{P_{\delta, \text{Phase 1}}^{\Lambda\text{CDM}}(z,k, \mathbf{p})+P_{\delta, \text{Phase 2}}^{\Lambda\text{CDM}}(z,k, \mathbf{p})}
\end{equation}
is within 2\% of the mean over 10 realizations while requiring only twice the computing time of a single run.

\begin{figure}[h!]
    \centering
    \includegraphics[width=1.0\linewidth]{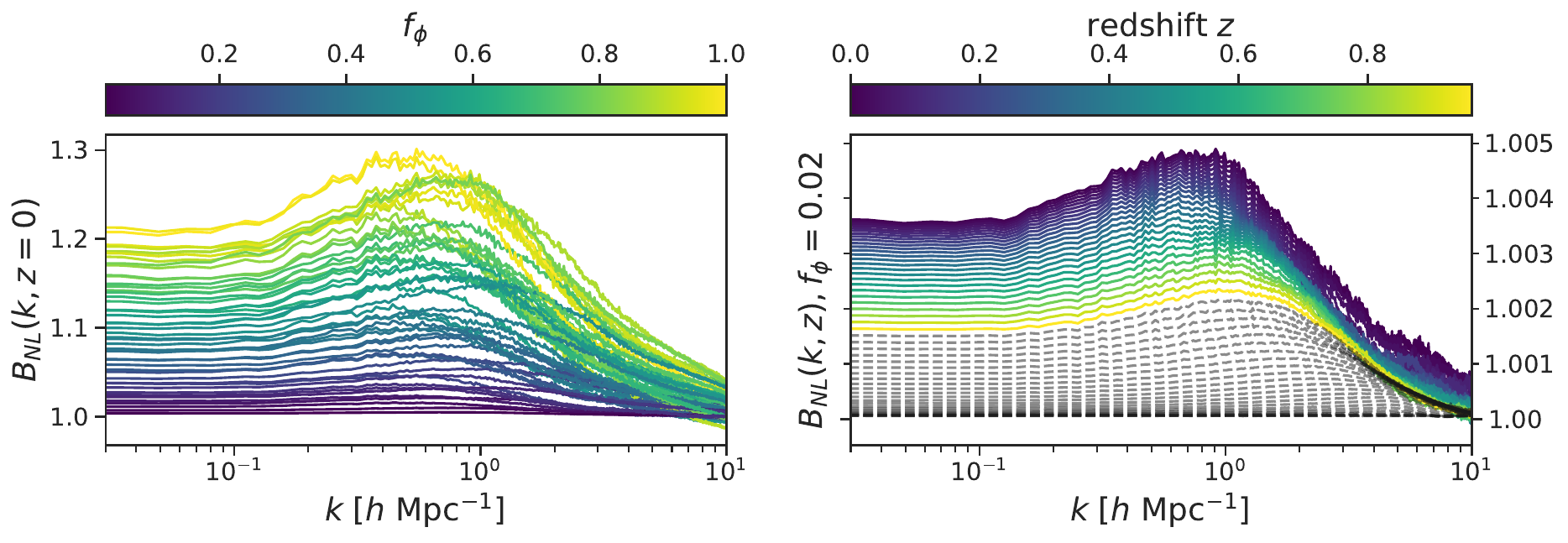}
  \caption{Visualization of the training boosts as a function of $k,z$ and $f_{\phi}$. Left: Plot of the training boost (defined in \ref{eq:Boost}) as a function of $k$ at $z=0$ for our training data. The colour describes variation in the parameter $f_\phi$. Right: Plot of the training boost as a function of $k$ and $z$ for our training data (at our lowest value of $f_\phi$, $f_\phi = 0.02$). The colour describes variation in redshift (for clarity, we don't display the redshifts for $z\geq 1$, opting for dotted black lines). This plot shows that $f_\phi$ is sampled densely enough close to the $\Lambda$CDM prediction ($B=1$).}
    \label{fig:Training_Boosts}
\end{figure}

The boosts were computed at 51 different redshifts in the interval $0 \leq z \leq 49$ linearly spaced in the scale factor $a$, and for $768$ logarithmically spaced wavevectors over the range $5 \times 10^{-2} \,h\,\text{Mpc}^{-1} \leq k \leq 10 \,h\,\text{Mpc}^{-1}$. The \HiCOLA{} power spectrum boosts have been validated to within 2.5\% of high-fidelity N-body simulations for $k \lesssim 1.2 \, h\,\text{Mpc}^{-1}$ \cite{Wright_2023} for some Cubic Galileon models -- the boost (and therefore the emulator) is at most reliable up to these scales. The validation boosts are shown in Figure \ref{fig:Training_Boosts}. These qualitatively reproduce the characteristic behaviour expected from Cubic Galileon models: a plateau at small $k$ consistent with the linear prediction of scale-independent growth, a bump on quasi-linear scales arising from the enhanced collapse of structures whose fifth force has not yet been suppressed, and a turn-over at higher k, where the Vainshtein screening mechanism causes the boost to tend towards the $\Lambda$CDM prediction. All these features become more pronounced with increasing $f_\phi$, reflecting the stronger modification to gravity as the scalar field contribution increases.

Finally, to reduce the noise arising from the finite simulation volume and resolution, the boost is smoothed (in wavevector $k$) prior to emulator training. Following standard practice (e.g. \cite{Ramachandra_2021, Mauland_2024}), we use a Savitzky–Golay filter \cite{savitzky1964smoothing} with window length 13, polynomial order 3, preventing the emulator from learning the noisy features of the data and improving its accuracy.

\section{Emulator Development}\label{sec:Emulator_development}

Having generated the suite of smoothed boosts described in Section \ref{sec:Emu_design}, we now construct an emulator that interpolates the boost $B(z,k,\mathbf{p})$ across the five-dimensional parameter space $\mathbf{p} = \{\Omega_{m0}, n_s, A_s, h, f_\phi\}$ at any redshift within the simulated ranges. Our emulator follows the approach pioneered by the \texttt{CosmicEmu} and Coyote Universe emulation program \cite{Heitmann_2009, 2010ApJ...713.1322L, 2014ApJ...780..111H}. For each simulation snapshot, we compress the boost into a small set of orthogonal basis functions and fit Gaussian processes to the corresponding weights. Predictions at intermediate redshifts are then obtained by interpolation between snapshots.

\subsection{Single-snapshot emulation}\label{sec:Emu_snapshot}

We first describe the emulator at a fixed simulation snapshot, where the boosts form a matrix of $N_{\rm sim} = 50$ simulated curves, each sampled on $N_k = 768$ logarithmically spaced wavevectors. Rather than emulating the boost at every $k$-bin independently, we compress this output onto a small set of orthogonal basis functions via a singular value decomposition (SVD). The smoothed boosts are centred and standardised, and we represent the boost as a truncated sum of basis vectors $\phi_m(k)$ and associated weights $w_m(\mathbf{p})$,
\begin{equation}\label{eq:pca}
    B(k,\mathbf{p}) \simeq \bar{B}(k) +
    \sum_{m=1}^{n_{\rm PC}} \phi_m(k)\, w_m(\mathbf{p}),
\end{equation}
where $\bar{B}(k)$ is the mean boost over the training set and the number of retained components $n_{\rm PC}$ is chosen to capture $99.9\%$ of the variance. Because the boost is a smooth, monotonic function of the parameters (see Figure \ref{fig:Training_Boosts} for the variation with $f_\phi$), only a handful of components are required (between 1 and 4 depending on the redshift). The orthogonality of the basis vectors $\phi_m(k)$ also allows the weights $w_m(\mathbf{p})$ to be treated independently for fitting Gaussian Process (GP) models. 

We have training inputs $\{\mathbf{p}_i\}$ and weights $\mathbf{w} = \{w_m(\mathbf{p}_i)\}$. Then for each $w$ in $\mathbf{w}$, the joint distribution of the training weights and the weight $w_*$ at a new parameter point $\mathbf{p}_*$ is taken to be Gaussian,
\begin{equation}\label{eq:gp_joint}
    p(w, w_*) = \mathcal{N}\!\left(
    \mathbf{0},
    \begin{bmatrix}
        \mathbf{K} & \mathbf{K}_*^{T} \\
        \mathbf{K}_* & \mathbf{K}_{**}
    \end{bmatrix}
    \right),
\end{equation}
where $\mathbf{K}$ is the covariance of the training points, with entries
$\mathbf{K}_{ij} = k(\mathbf{p}_i, \mathbf{p}_j)$ given by the kernel
evaluated on each pair of training inputs. The term $\mathbf{K}_*$ holds the
covariances between the test point and the training points, with entries
$k(\mathbf{p}_*, \mathbf{p}_i)$. The term
$\mathbf{K}_{**} = k(\mathbf{p}_*, \mathbf{p}_*)$ is the prior variance of
the test point.

Conditioning on the training data yields the predictive mean and variance at $\mathbf{p}_*$,

\begin{eqnarray}\label{eq:gp_predict}
    \mu(w_*) &=& \mathbf{K}_* \, \mathbf{K}^{-1} \mathbf{w}, \\
    \sigma^2(w_*) &=& \mathbf{K}_{**} - \mathbf{K}_* \, \mathbf{K}^{-1}
    \mathbf{K}_*^{T},
\end{eqnarray}

The predictive mean $\mu(w_*)$ is a weighted combination of the training weights, with the weighting set by how strongly the test point correlates with each training point through the kernel. The predictive variance $\sigma^2(w_*)$ is smallest where the test point lies close to the training nodes and grows in sparsely sampled regions of parameter space. 

The kernel function $k(\mathbf{p}, \mathbf{p}')$ encodes how smoothly the weights vary across parameter space. We adopt a squared-exponential kernel for each weight, so that the correlation between two parameter points decays smoothly with their separation. Each kernel has a separate correlation length for each of the five input dimensions corresponding to the cosmology parameters, and a small nugget term:

\begin{equation}
    k(\mathbf{p}, \mathbf{p}') = \boldsymbol{\tau}_{\text{SE}}^2 \text{exp}\left(-\frac{||\mathbf{p} - \mathbf{p}'||^2}{2\boldsymbol{l}^2}\right) + \boldsymbol{\tau}_{\text{noise}}^2\mathbb{I}
\end{equation}
where $\mathbb{I}$ is the identity matrix. This form is appropriate because the boost responds smoothly to all five parameters across the ranges in eq.~\eqref{eq:param_ranges}. The kernel
hyperparameters are inferred within the Bayesian framework of \cite{Higdon2008} as implemented in \texttt{SEPIA} \cite{james_gattiker_2020_4048801}, sampling the posterior with a Markov Chain Monte Carlo of $1000$ steps. The trained GPs are then combined with the basis vectors of eq.~\eqref{eq:pca} to predict the boost and its uncertainty at any parameter point within the snapshot. 

\subsection{Redshift interpolation}\label{sec:Emu_redshift}

The procedure above is repeated independently at each of the $51$ simulation snapshots, giving a series of 51 separate emulators spanning $0 \leq z \leq 49$. 
We treat redshift separately, opting to not include it as a sixth dimension in the Latin hypercube design, or compress it via principal component analysis (like $k$). This has a practical advantage: every simulation outputs the boost on the same fixed set of redshifts, so the training data shares a common time grid across all nodes, instead of each node sampling a different redshift as would follow from folding $z$ into the experimental design. It is also well motivated, since the boost varies smoothly with redshift. Hence, a full GP in that direction would be unnecessary.

To evaluate the boost at an arbitrary redshift $z$ between two stored snapshots $z_1$ and $z_2$, we deploy the single-redshift emulators at the two nearest snapshots and linearly interpolate between them,
\begin{equation}\label{eq:z_interp}
    B(z,k,\mathbf{p}) = B(z_2,k,\mathbf{p}) +
    \big[B(z_1,k,\mathbf{p}) - B(z_2,k,\mathbf{p})\big]
    \frac{z - z_2}{z_1 - z_2},
\end{equation}
propagating the per-snapshot emulator errors through the same linear weighting. Validations of the redshift interpolation are shown in Section \ref{sec:Emu_results}, where the emulator is deployed at redshifts not used in the emulator training. 

\section{Modelling and Forecasting Methods}\label{sec:Inference_methods}
Having constructed the Cubic Galileon emulator, we now turn to its principal application in this work, namely as a forecasting tool for constraints on Cubic Galileon gravity with upcoming photometric survey data. The emulator's speed makes it possible to model nonlinear modified gravity predictions directly within a Bayesian inference pipeline, a task that would be computationally prohibitive with simulations alone. In this section we describe the inference pipeline used in Section \ref{sec:forecasts}, where we produce forecasts for the constraints LSST Y1 will place on the modified gravity parameter $f_\phi$, and additionally test the emulator's accuracy for a range of scale cuts.

We consider the $3\times2$pt datavector, which has become a standard choice for photometric galaxy surveys \cite{descollaboration2026darkenergysurveyyear, Heymans_2021}. It combines three sets of two-point correlation functions: the auto-correlation of the galaxy density field (galaxy clustering), the auto-correlation of the weak lensing shear field (cosmic shear), and the cross-correlation between the two (galaxy-galaxy lensing). The 3$\times$2pt datavector is particularly well suited to tests of gravity. Galaxy clustering is sensitive to the Newtonian potential ($\Phi$), so to modifications of the Poisson equation, which can be parametrized through the function $\mu$:
\begin{equation}
    k^2 \Phi = -4\pi G_N \mu(a,k) a^2 \rho_m \delta
\end{equation}
while lensing responds to modifications to the equation governing the Weyl potential ($\Phi_+$) and therefore to $\Sigma$:
\begin{equation}
    k^2 \Phi_+ = -4\pi G_N \Sigma(a,k) a^2 \rho_m \delta
\end{equation}
where $G_N$ is Newton’s constant, $\rho_m$ is the background density of matter and $\delta$ is the comoving matter density contrast. $\mu$ and $\Sigma$ are dimensionless phenomenological functions equal to unity in $\Lambda$CDM. Because $\Sigma$ affects the photon geodesics only, it will affect the equations describing lensing, but not those describing galaxy clustering. 
These equations are true under the quasi-static approximation \cite{Silvestri_2013}, neglecting anisotropic shear, and on linear scales. We can extend these definitions on nonlinear scales by re-defining the functions $\mu$ and $\Sigma$ accordingly (e.g. \cite{Thomas_2020}), but a full treatment is beyond the scope of this paper. Our forecasts make the simplifying assumption that $\Sigma$ takes its linear form throughout, but we note here that this is not guaranteed to be a valid approximation for the weak lensing angular power spectrum on nonlinear scales, and should be revisited in a realistic analysis. Additionally, the approximation does not affect the internal consistency of our bias tests, as the same treatment of $\Sigma$ enters both the mock datavectors and the model. We expect that modifications to posterior biases will therefore be small.

Additionally, this $3\times2$pt tracer combination allows a better handle over systematics. For example, the combination of lensing and clustering amplitudes alleviates the degeneracy between cosmological parameters and linear galaxy bias. In addition, the full 3$\times$2pt datavector contains significant information from the non-linear regime, where screening mechanisms are expected to operate and where departures from GR acquire scale dependence. Modelling this datavector, however, requires a good handle over observational systematics such as intrinsic alignments, photometric redshift calibration, galaxy bias and baryonic feedback, especially on nonlinear scales \cite{DESCSRD2021lsst}.

In what follows, we describe the construction of the simulated data vector, our modelling choices and scale cuts, and the likelihood and sampling setup, before forecasting the resulting constraints on the modified gravity parameter $f_\phi$ and the cosmological parameters.
\newline

\noindent \textbf{\textit{Priors and Parameter Values.}} Table \ref{table:p_fid_and_priors} summarizes the fiducial values and priors used in our 3$\times$2pt analysis for cosmology ($\mathbf{p}_{\mathbf{co}}$), modified gravity ($f_{\phi}$), and nuisance ($\mathbf{p}_{\mathbf{nu}}$) parameters used in our analysis. Nuisance parameters ($b^i$, the linear galaxy bias parameters for each lens redshift bin $i \in \{1, ..., n_{g}\}$, where $n_{g}$ is the total number of lens bins) are taken from LSST Y1 forecasts \citep{DESCSRD2021lsst} (with priors from \cite{DES_Y3_results}), and flat priors are imposed on cosmological parameters due to emulator limitations (as described in eq. \ref{eq:param_ranges}). The only informative prior is the joint Gaussian prior on $\omega_b$ and $n_s$. We introduce CMB-informed priors, without adopting the full {\it Planck} likelihood. Our prior comes from the marginalized posterior distributions for the {\it Planck} TT,TE,EE+lowE+lensing data using the late-time DE parametrization (outlined in Equations 56a and 56b in \cite{2020}, see Fig. 31). 
These priors are motivated by the weak constraints that cosmic shear alone places on 
$n_s$ and $\omega_b$ \cite{Hall_2021, tsedrik2024stageivcosmicshearmodified}, and by the expectation that any realistic MG analysis will incorporate CMB information to break degeneracies. The choice of informative prior is not unique: CMB priors are also commonly imposed on $A_s$, or alternatively less constraining BBN-only priors on $\omega_b$ can be adopted\footnote{Relaxing these priors may redistribute the parameter biases in the forecasts, shifting them from the nuisance parameters onto $n_s$, $\omega_b$ and other degenerate cosmological parameters like $A_s$ (we have tested that this can indeed be the case, though the effect is relatively minor for the case of the more conservative baryonic scale cuts introduced below, and with some priors on $\omega_b$).}. Since our aim here is to demonstrate a preliminary use case for the emulator rather than to characterise biases and evaluate degeneracies for a range of choices of priors and scale cuts, we leave further investigations along this line of inquiry to future work.
The procedure to extract this prior is as follows. We marginalise the posterior over all parameters except $\omega_b$ and $n_s$, and then extract a joint covariance for these two parameters under the assumption that their joint posterior is Gaussian. Finally, we shift the joint mean to match the fiducial values of $\omega_b$ and $n_s$.
\newline
 
\noindent \textbf{\textit{Simulated data.}} We will compute the correlation functions between two observable fields: cosmic shear and galaxy clustering. Because LSST is a photometric survey, galaxies are grouped into tomographic redshift bins based on their photometric redshift estimates ($n_s$ bins for the source galaxies sample, and $n_g$ bins for lens galaxies sample), and statistical correlations are then computed between the observable fields within these bins. Our 3$\times$2pt datavector is the combination of three sets of angular power spectra (i.e. 2-point statistics) $\textbf{D} = \{C^{ii}_{\delta\delta}(\ell),C^{ij}_{\kappa\delta}(\ell),C^{ij}_{\kappa\kappa}(\ell)\}$, where: 

\begin{itemize}
    \item $C^{ii}_{\delta\delta}(\ell)$ describes the correlations in the galaxy density contrast from galaxies in $n_{g}$ different lens redshift bins, with $i \in \{1, ..., n_{g}\}$, in projected Fourier mode $\ell$. We only include the autocorrelation in redshift bins.
    
    \item $C^{ij}_{\kappa\kappa}(\ell)$ describes the correlations in the galaxy shear in $n_{s}$ different source redshift bins, with $i,j \in \{1, ..., n_{s}\}$, in projected Fourier mode $\ell$.
    
    \item $C^{ij}_{\kappa\delta}(\ell)$ describes the cross-correlation of galaxy shear in the $i^{th}$ source galaxy bin with the galaxy density contrast of the $j^{th}$ foreground lens galaxy bin, with $i \in \{1, ..., n_{s}\}$ and $j \in \{1, ..., n_{g}\}$, in projected Fourier mode $\ell$.
\end{itemize}

\begin{figure}
    \centering
    \includegraphics[width=1.0\linewidth]{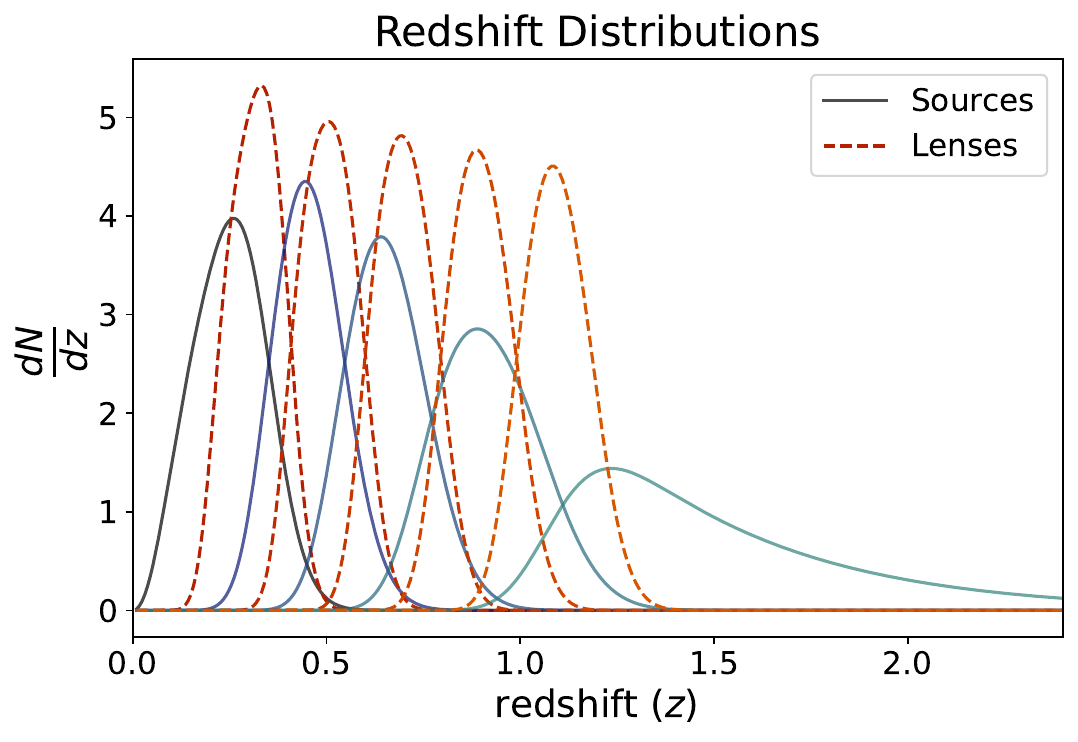}
    \caption{The normalized galaxy number density for the source (solid) and lens (dashed) distributions for the 5 tomographic redshift bins from low-z (darkest) to high-z (lightest).}
    \label{fig:z_dist}
\end{figure}

We compute the angular power spectrum by using the Cubic Galileon emulator boost in combination with the Core Cosmology Library\footnote{\url{https://github.com/LSSTDESC/CCL}} (\texttt{CCL}) \citep{Chisari_2019b}, which uses the halofit transformation of \cite{Takahashi_2012} to estimate the nonlinear $\Lambda$CDM 3D matter power spectrum. The angular power spectrum between redshift bin $i$ of observable $A$ and redshift bin $j$ of observable $B$ at multipole $\ell$, under the Limber approximation \citep{Limber_approx} is:
\begin{align}\label{eq:C_ell_general}
    C^{ij}_{AB}(\ell) = \int d\chi \frac{q^i_A\left(\chi \right) q^j_B\left(\chi \right)}{\chi^2} P_{\text{CuGal}}\left(\frac{\ell+\frac{1}{2}}{\chi}, z(\chi)\right).
\end{align}
The tracers $\text{AB}$ are either cosmic shear ($\kappa \kappa$), galaxy-galaxy clustering ($\delta \delta$, labeled GC) or their cross correlation ($\delta \kappa$, labeled XC), $\chi$ is the comoving radial distance and $i$ and $j$ denote redshift bin indices. $q^i_\kappa(\chi)$ and $q^i_\delta(\chi)$ are line-of-sight kernels, which weigh the power spectrum across all distances along the line of sight. These are given by:
\begin{eqnarray}\label{eq:kernels}
 q^i_\kappa (\chi) &=& \frac{3H_0^2 \Omega_m\chi \Sigma_{\text{CuGal}}(z(\chi))}{2a(\chi)} \int_{\chi}^{\chi_h} d\chi^\prime \left(\frac{\chi^\prime - \chi}{\chi'}\right) n^i_s(z(\chi^\prime)) \frac{d z}{d\chi^\prime} , \nonumber \\
 q^i_\delta (\chi) &=& b^i\  n^i_g(z(\chi)) \frac{d z}{d\chi}\ .
\end{eqnarray}
where $a$ is the scale factor and  $b^i$ is a constant galaxy bias for each lens redshift bin (that can in theory be scale- as well as redshift-dependent). The function $\Sigma_{\text{CuGal}}(z)$ describes deviations from GR in the lensing potential, and $\Sigma_{\text{CuGal}}(z) = \mu_{\text{CuGal}}(z)$ \cite{Peirone_2019,Frusciante_2020,Noller_2021}. Section 2.2 of \cite{Wright_2023} confirms that this equality holds on nonlinear scales. As mentioned above, in this paper we make the further assumption that $\Sigma=\mu_{\rm linear}$.
$n^i_{s, g}(z)$ denote the redshift distributions of the different source/lens redshift bins, normalized to unity (see Figure \ref{fig:z_dist}). Following \cite{DESCSRD2021lsst}, we assume a parametric redshift distribution for our overall underlying `true' source and lens distributions:
\begin{align}
    \frac{\text{d}n_{s,g}}{\text{d}z}\propto z^2\text{exp}[-(z/z_0)^\alpha].
\end{align}
The source sample is separated in 5 bins with equal number counts, and with $\{z_0,\alpha\} = \{0.13, 0.78\}$ and with photo-$z$ scatter $\sigma_z = 0.05(1 + z)$. The lens sample is separated in 5 bins separated by $0.2$ in photo-$z$ in the redshift interval $0.2 \leq z \leq 1.2$, with $\{z_0,\alpha\} = \{0.26,0.94\}$ and with photo-$z$ scatter $\sigma_z = 0.03(1 + z)$. Our model for this follows \cite{DESCSRD2021lsst} in approximating the effect of photo-z uncertainty without simulating a more complicated treatment of photo-z errors.

To ensure our forecasts are not biased by the specific nonlinear model assumptions built into our $\Lambda$CDM matter power spectrum predictions, we construct all power spectra consistently. Both for the simulated datavector and for the model predictions, rather than using the nonlinear Cubic Galileon power spectrum directly (from the simulations), we compute the Cubic Galileon power spectrum as the product of its boost and a $\Lambda$CDM model baseline:
\begin{align}
P_{\text{CuGal}} = B_{\text{CuGal}} \times P_{\Lambda\text{CDM, CCL}}.
\end{align}
This approach isolates the MG signal and avoids importing modelling differences from the GR calculation between \texttt{CCL} (halofit) and the N-body codes (both \HiCOLA{} and \texttt{EFT-RAMSES}) into our MG forecasts. The boost $B$ is either the one predicted from the emulator or from a \HiCOLA{} or \texttt{EFT-RAMSES} simulation. The extrapolation procedure for the latter is described in Section \ref{sec:forecasts_results_nbody}. For the former, we extrapolate beyond $z=12$ to higher $z$ by assuming that the power spectrum is close to GR (so we set $B(k,z\geq12) = 1$; this is a good approximation because the MG boost is always within 0.3\% of 1 at those redshifts). Additionally, if $k\geq5 \, h\,\text{Mpc}^{-1}$, we use a power-law extrapolation for the boost: $B(k)=Ak^{-p} +1$, with $p\geq0$ always (for a similar approach, see e.g., \cite{tsedrik2024stageivcosmicshearmodified}). 
 
Since the $\chi^2$-based scale cuts of Section \ref{sec:forecasts} require an input datavector, before applying scale cuts we model $20$ logarithmically spaced $\ell$ bins over $20 \leq \ell \leq 15000$, following \cite{DESCSRD2021lsst} to match the DESC Science Requirements Document datavector. In practice these high-$\ell$ scales do not enter our analysis, as the baryonic cuts restrict us to $\ell$ of a few hundred and our less conservative cuts to $\leq1500$.

For galaxy clustering, we only consider the lens bins auto-correlations $C^{ii}_{\delta_g\delta_g}(\ell)$. Cross-correlations between lens bins do not contribute significantly to the signal, which rather than carrying genuine cosmological information is dominated by photo-z outlier systematics \cite{Schneider_2006}. For galaxy-galaxy lensing correlations, we only consider lens and source bin combinations where the source bins are at a higher redshift than the lens bins (allowing for at most a 10\% overlap). The selected bin pairs are $( n^i_s, n^j_g) = \{(3,1), (4,1), (5,1), (4,2), (5,2), (5,3), (5,4)\}$ \citep{DESCSRD2021lsst}.
\newline

\noindent \textbf{\textit{Baryonic scale cuts.}} While all the power spectra presented in this work are dark matter only (DMO), a real datavector will be affected by baryonic physics, in particular by the impact of galaxy formation processes (e.g., AGN feedback, cooling, star formation) on the distribution of matter \cite{Chisari_2019}. To produce more realistic and conservative forecasts, we account for this lack of baryonic physics modelling in our angular power spectra by applying scale cuts in $\ell$.
We adopt this approach because baryonic modelling in modified gravity remains relatively poorly understood. Baryonification schemes are available\footnote{\url{https://github.com/sambit-giri/BCemu}} and have been applied in modified gravity forecasts, typically as an additional baryonic boost to DMO power spectra \cite{srinivasan2026cosmologicalgravityscalesv}. Investigation of the degeneracies between the baryonification parameters and the modified gravity parameters are being undertaken in order to ensure these tools can be used in a realistic MG analysis.  \cite{Mead_2016} first showed that, in a halo model context, MG and baryons can be modelled independently, and the limitations of this approximation have been tested for specific modified gravity theories (see e.g. \cite{Arnold_2019}). Still, scale cuts are likely to be adopted in practice for early LSST MG analyses. The use of less conservative approaches is possible (for example retaining smaller scales via the BNT transform, see \cite{Taylor_2021, Srinivasan_2025}) and applying it when forecasting Cubic Galileon gravity constraints with $3\times2$pt data is a direction of future work.

We therefore describe now our procedure for cutting scales from the 3$\times$2pt analysis that are sensitive to baryonic physics. There will be a maximum wavevector beyond which nonlinear bias results in significant deviations in the power spectrum from the linear bias, which we call $k_{\text{max}}$. We fix this scale to $k_{\text{max}} = 0.2 h\,\text{Mpc}^{-1}$, a slightly more conservative value than the one suggested in \cite{DESCSRD2021lsst}. We note here that the linear galaxy bias model is expected to deliver unbiased (sub-percent) constraints only for $k \lesssim  0.1\, h/{\rm Mpc}$ \cite{truttero2026balancingbiasbaryonsscale}. We retain $k_{\max}=0.2\,h\,{\rm Mpc}^{-1}$ as a compromise between this limit and the value adopted in the DESC SRD.  Unbiased recovery at
this $k_{\max}$ implies unbiased recovery for any more conservative choice, such as $k_{\max}=0.1\,h\,{\rm Mpc}^{-1}$. We find $\ell_{\text{max}}$ using the following approximation (see e.g. \cite{grandis_2024}):
\begin{align}\label{eq:clustering_cut}
    \ell_{\text{max}} = k_{\text{max}}\chi(\langle z \rangle) - 0.5
\end{align}
where $\langle z \rangle$ the mean redshift of the relevant lens bin.

We use this method to implement scale cuts for $C^{ij}_{\delta\delta}(\ell)$ and $C^{ij}_{\delta\kappa}(\ell)$, but not for $C^{ij}_{\kappa\kappa}(\ell)$, given Equation \eqref{eq:clustering_cut} only works for narrow kernels and the lensing kernel is broad. For the latter, we will instead use a method based on \cite{Doux_2022}. This involves comparing two synthetic, noiseless datavectors for a DMO and baryonic-rescaled $\Lambda$CDM universe computed at a fiducial cosmology:
$$
    P_{\text{baryonic}}(\mathbf{p}^{\text{fid}},k,z)= \frac{P^{\text{sim}}_{\text{OWLS}}(k,z)}{P^{\text{sim}}_{\text{DMO}}(k,z)} \times P_{\text{DMO}}(\mathbf{p}^{\text{fid}},k,z) 
$$
where the power spectra come from the OWLS simulations  \citep{van_Daalen_2011}, $P^{\text{sim}}_{\text{DMO}}(k,z)$ for DMO and $P^{\text{sim}}_{\text{OWLS}}(k,z)$ with AGN feedback only, matching the baryonic scenario adopted in the DES Y3 analysis \citep{Abbott_2023}. After computing the shear angular power spectrum datavectors using Equation \eqref{eq:C_ell_general}, we compute:
\begin{equation}\label{eq:chi2criterion}
    \Delta \chi^2 = \left(\mathbf{D}_{\text{baryonic}}^{\Lambda\text{CDM}} - \mathbf{D}_{\text{DMO}}^{\Lambda\text{CDM}}\right)^T \mathbf{C}^{-1} \left(\mathbf{D}_{\text{baryonic}}^{\Lambda\text{CDM}} - \mathbf{D}_{\text{DMO}}^{\Lambda\text{CDM}}\right)    
\end{equation}
where $\mathbf{C}$ is the data covariance matrix described further below. We then identify the point that contributes the most to $\Delta \chi^2$, remove it and repeat the process until $\Delta \chi^2 \leq 1$. 

The criteria of Eq. \ref{eq:chi2criterion}, when enforced on the full $3\times2$pt datavector up to $\ell = 15000$, retains counterintuitively the two largest $\ell$ bins of the $C^{11}(\ell)$ datavector for galaxy-galaxy lensing. While we expect the greatest difference in nonlinear vs linear modelling at the smallest scales, this condition is allowed by the criteria due to the role of the covariance matrix, potentially in combination with inaccurate modelling of the power spectrum for the largest $\ell$ values. We therefore decide to additionally cut these points in our analysis; we enforce this by removing all $\ell \geq \ell_{\text{max }\Delta \chi^2}$ during the iteration.
\newline

\noindent \textbf{\textit{Linear scale cuts.}} For comparison with the expected constraints of a forecast based on linear-only power spectrum modelling, we will implement the following scale cuts on the 3$\times$2pt datavector based on \cite{Doux_2022}:
\begin{equation*}
    \Delta \chi^2 = \left( \mathbf{D}^{\Lambda\text{CDM}}_{\mathrm{NL}} - \mathbf{D}_{\mathrm{lin}}^{\Lambda\text{CDM}} \right)^{T} 
    \mathbf{C}^{-1} 
    \left( \mathbf{D}_{\mathrm{NL}}^{\Lambda\text{CDM}} - \mathbf{D}_{\mathrm{lin}}^{\Lambda\text{CDM}} \right) \leq 1.
    \label{eq:deltachi2}
\end{equation*}
We do this by finding the point in the full 3$\times$2pt datavector that contributes the most to $\Delta \chi^2$, removing it, and repeating the process until $\Delta \chi^2 \leq 1$.
\newline

\noindent\textit{\textbf{Likelihood formalism.}} We describe here the general method for computing the likelihood in our simulated analysis. We first compute our mock data vector for some cosmological parameters (labeled `$\text{fid}$' -- unless otherwise specified, see Table \ref{table:p_fid_and_priors} for the fiducial values) and some choice of $f_\phi$:

$$\mathbf{D} = \{C^{ii,\text{NL}}_{\delta_g\delta_g}(\ell , \mathbf{p}_{\text{co}}^{\text{fid}},\mathbf{p}_{\text{nu}}^{\text{fid}},f_\phi),C^{ij,\text{NL}}_{\kappa\delta_g}(\ell , \mathbf{p}_{\text{co}}^{\text{fid}},\mathbf{p}_{\text{nu}}^{\text{fid}},f_\phi),C^{ij,\text{NL}}_{\kappa\kappa}(\ell , \mathbf{p}_{\text{co}}^{\text{fid}},\mathbf{p}_{\text{nu}}^{\text{fid}},f_\phi)\}.$$

We do this by using a 3D matter power spectrum interpolated from either \HiCOLA{} (studied in Sec.~\ref{sec:forecasts_results}) or \texttt{EFT-RAMSES} simulations (studied in Sec.~\ref{sec:forecasts_results_nbody}). We then find our model datavector as a function of model parameters $\mathbf{p}$:

$$\mathbf{M}(\mathbf{p}) = \{C^{ii,\text{NL}}_{\delta_g\delta_g}(\ell,\mathbf{p}),C^{ij,\text{NL}}_{\kappa\delta_g}(\ell,\mathbf{p}),C^{ij,\text{NL}}_{\kappa\kappa}(\ell,\mathbf{p})\}$$
using the 3D matter power spectrum prediction from the emulator. For chains that implement linear scale cuts, we use the 3D linear matter power spectrum instead. We then model the likelihood as Gaussian in the datavector and as a function of the parameters $\mathbf{p}$:

\begin{align}\label{eq:likelihood}
    \begin{split}
        \text{ln} \mathcal{L} (\mathbf{D}|\mathbf{p})\propto &-\frac{1}{2} \left[ (\mathbf{D} - \mathbf{M}(\mathbf{p}))^T \mathbf{C}^{-1} (\mathbf{D} - \mathbf{M}(\mathbf{p}))\right]
    \end{split}
\end{align}
where $\textbf{C}$ is the full non-Gaussian Y1 3$\times$2pt covariance\footnote{\href{https://github.com/CosmoLike/DESC\_SRD}{https://github.com/CosmoLike/DESC\_SRD}} described in \cite{DESCSRD2021lsst} (created following the numerical prescription from \cite{Krause_2017}). While this covariance is pre-computed at different cosmological parameter values ($\mathbf{p}^{\text{fid, SRD}}_{\text{co}} = \{\Omega_m = 0.3156, \sigma_8 = 0.831,n_s =0.9645,\Omega_b =0.0492,h =0.6727\}$) \citep{DESCSRD2021lsst} to our fiducial ones (\ref{table:p_fid_and_priors}), this should not significantly affect results (see \cite{Kodwani_2019}).

We sample the posterior defined by Equation \eqref{eq:likelihood} and the priors of  Table \ref{table:p_fid_and_priors} over the full cosmological, modified gravity, and  nuisance parameter space in Section \ref{sec:forecasts} below. The machinery described in this section allows us to perform  a simulated analysis, by generating $\mathbf{D}$ from an independent \HiCOLA{} or \texttt{EFT-RAMSES} simulation,  while modelling $\mathbf{M}(\mathbf{p})$ with our emulator. Any offset between the recovered and input parameters directly measures the bias induced by emulator at the level of cosmological inference. At the same time, the resulting posteriors give an estimate of the constraining power of an LSST Y1 $3\times2$pt analysis on $f_\phi$. In the next section we present the results of these forecasts, alongside direct comparisons of  the emulated boosts against validation \HiCOLA{} and \texttt{EFT-RAMSES} simulations.
\newline

\section{Results}\label{sec:Emu_results}

The emulator’s accuracy is limited by a range of factors. Firstly, approximations in the underlying simulations used to produce the training set introduce error: the finite simulation box size affects large-scale accuracy, and in order to reduce computational cost we use lower-resolution COLA simulations instead of full N-body, but in doing so decrease small-scale accuracy. Secondly, errors arise during emulator development due to the limited number of training samples, and the methods used for training data selection and interpolation.

To assess these effects, we conduct two kinds of tests. In the first, we compare the emulator's predictions with those from \HiCOLA{} simulations that were not used in training. We do so by comparing the boost predictions directly (in Section \ref{sec:COLA_validation} below, where we compare with results from two additional sets of respectively 25 and 50 simulations), or by running simulated forecasts for 3$\times$2pt LSST Y1 constraints based on \HiCOLA{} simulations (in Section \ref{sec:forecasts_results}). This helps us measure the accuracy of the emulator across parameter space and its effectiveness as a forecasting tool for LSST Y1, relative to \HiCOLA{} simulations of the same kind it was trained on. In the second test, we compare the emulator’s performance against a higher resolution \texttt{EFT-RAMSES} N-body simulation to evaluate its overall effectiveness, by running simulated forecasts for 3$\times$2pt LSST Y1 constraints (in Section \ref{sec:forecasts_results_nbody} below)\footnote{For a direct comparison of the boosts and of the $3\times2$pt LSST Y1 datavectors between \texttt{EFT-RAMSES}, \HiCOLA{} and the emulator, see Appendix A.1 in Cubic\_Galileon\_Emu\_appendices\_github.pdf in \href{https://github.com/CarolaZano/CubicGalileon_Forecasts}{github.com/CarolaZano/CubicGalileon_Forecasts}}. This allows us to assess the performance of the emulator by separately investigating errors from both the COLA simulations and the emulator construction. 

\subsection{Emulator Accuracy and Validation: comparison with \HiCOLA}\label{sec:COLA_validation}

Here we introduce the \HiCOLA{} simulated datasets used for validation.

\paragraph{Dataset 1: varying $f_\phi$ only} We fixed the cosmological parameters to the fiducial values in Table \ref{table:p_fid_and_priors} and sampled $f_\phi$ uniformly between 0 and 1, choosing values that lie between those used in the emulator's training set (in red, compared with the training-set values in blue). We also increased the number of redshift timesteps to 100, both to improve the accuracy of the simulations and to assess the emulator's accuracy at redshifts other than those it was trained on.

\begin{table}
\centering
 \begin{tabular}{m{1.5em} m{4.1em} m{6em} m{24em}} 
 \hline\hline
 &Parameter & Fiducial Value & Priors\\ [0.5ex] 
 \hline\hline
 & $\Omega_m$ & 0.321 & $\mathcal{U}(0.275, 0.33)$ \\ 
 & $h$ & 0.6688 & $\mathcal{U}(0.61, 0.73)$ \\
 $\mathbf{p}_{\text{co}}$ & $A_s$ & $2.092\times 10^{-9}$ & $\mathcal{U}(1.45\times 10^{-9}, 3.3\times 10^{-9})$ \\
 & $\omega_b$ & 0.022383 &  \multirow{2}{*}{\resizebox{1.0\hsize}{!}{$ \mathcal{N}\left(\left[\begin{array}{cc}
    0.022383    \\
    0.9626 
 \end{array}\right],\left[\begin{array}{cc}
    2.0381\times10^{-5} & 3.6116\times10^{-7}   \\
    3.6116\times10^{-7} & 2.5265\times10^{-8}
 \end{array}\right]\right)$}}\vspace{0.4em}\\
 & $n_s$ & 0.9626 & \\
 \hline 
 & $b_1$ & 1.562362 & $\mathcal{U}(0.8, 3.0)$\\
 & $b_2$ & 1.732963 & $\mathcal{U}(0.8, 3.0)$ \\
 $\mathbf{p}_{\text{nu}}$ & $b_3$ & 1.913252 & $\mathcal{U}(0.8, 3.0)$ \\
 & $b_4$ & 2.100644 & $\mathcal{U}(0.8, 3.0)$ \\
 & $b_5$ & 2.293210 & $\mathcal{U}(0.8, 3.0)$ \\
 \hline 
  & $f_{\phi}$ & 1.0 & $\mathcal{U}(0.0, 1.0)$ \\
 \hline
 \hline
 \end{tabular}
 \caption{Priors and fiducial (i.e. simulated universe) parameters ($\mathbf{p}^{\text{fid}}$) for the cosmological, nuisance and modified gravity parameters in our simulated analysis. We quote the lower and upper limits of flat priors and the mean and standard deviation of Gaussian priors. The priors for the standard scale cuts (``Baseline Prior distributions'') match those in \cite{DES_Y3_results}.}
\label{table:p_fid_and_priors}
\end{table}

\begin{figure}[h!]
  \centering
  \makebox[\textwidth][c]{\includegraphics[width=1.0\textwidth, angle =0]{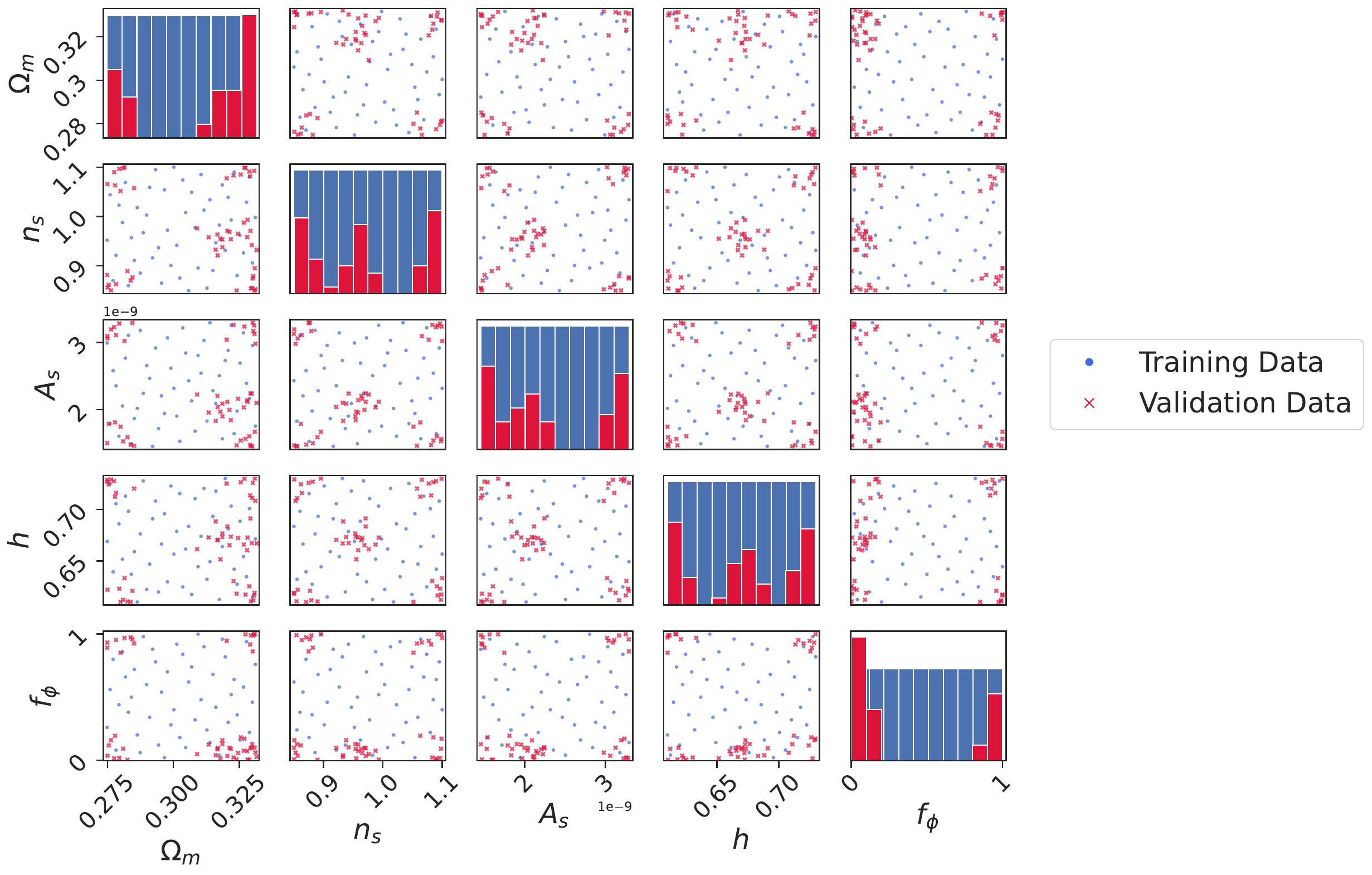}}%
  \caption{50 parameter combinations used for our second validation set (in red) compared to the 50 parameter combinations used in the emulator development (in blue).}
  \label{fig:Validation_param_set}
\end{figure}

\paragraph{Dataset 2: sampling the centre and edges}
The second validation dataset samples 50 points in parameter space, chosen to give representative coverage of both the centre and the edges of the parameter hypervolume. This allows us to assess the performance both in the region most important for parameter estimation (near the fiducial parameter values) and in the areas we expect the emulator will perform the worst (the edges). The hyperparameter space spanned by these points (in red) as compared to the training dataset (in blue) is displayed in Figure \ref{fig:Validation_param_set}.

We now assess the accuracy of the emulator by comparing its boost predictions to the boost from \HiCOLA{} simulations across the two validation datasets.

\begin{figure}[h!]
    \centering
    \includegraphics[width=1.0\linewidth]{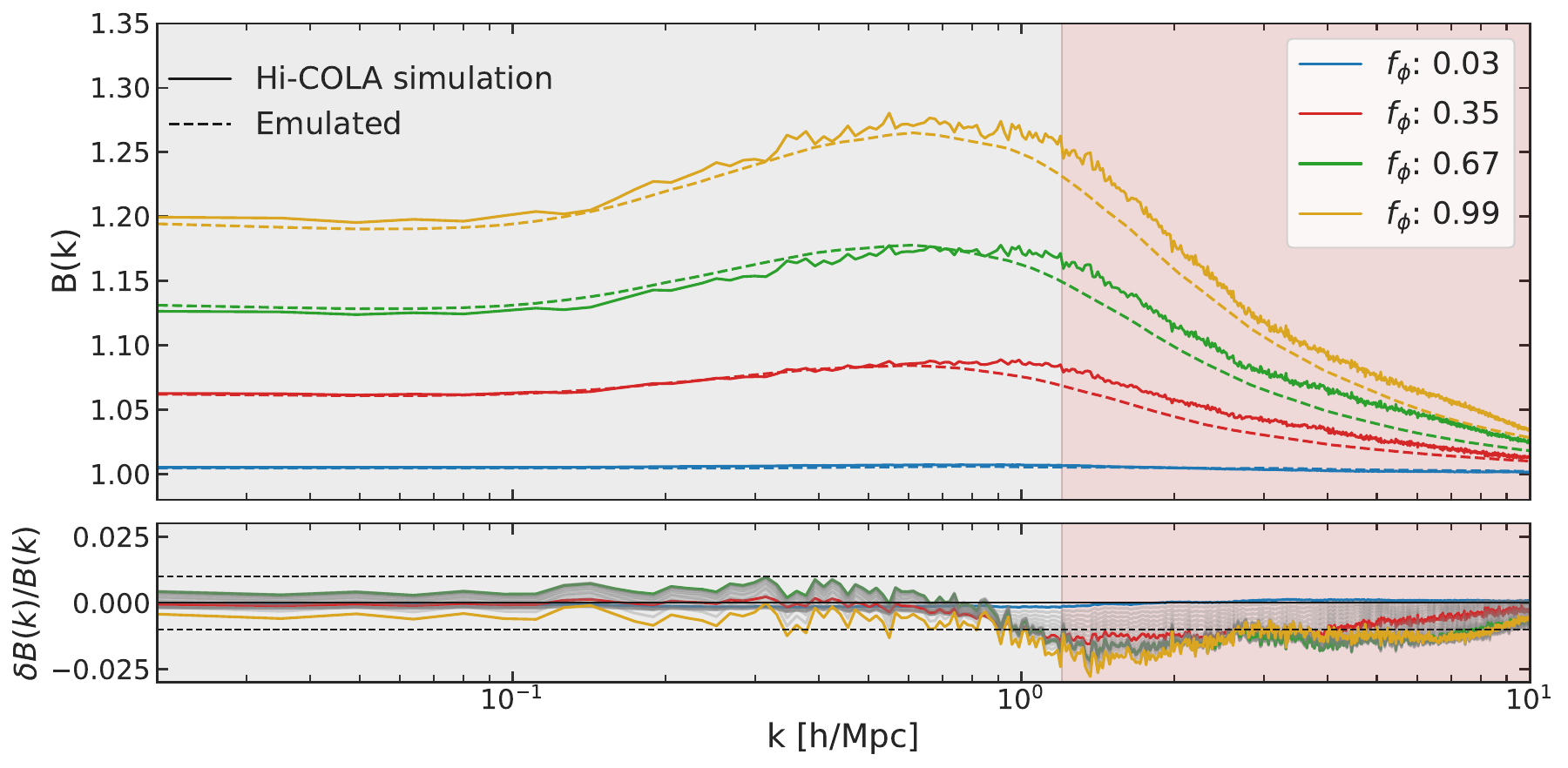}
    \caption{Top panel: Boosts predicted by the Cubic Galileon emulator (dashed lines) and the \HiCOLA{} boosts (solid lines) at $z=0$ for the validation dataset 1. Bottom panel: relative error between the emulator prediction and the \HiCOLA{} boosts at $z=0$ for the same validation set. In gray, we also show boost residuals at the remaining $f_\phi$ values for which there was validation data. The emulator is trained on a set smoothed with a Savitzky–Golay filter (window length 13, polynomial order 3), while the validation set was not smoothed. Errors remain $\sim 1\%$ for $k\lesssim 0.9 h\,\text{Mpc}^{-1}$ and under $2\%$ for $k\lesssim 1.2 h\,\text{Mpc}^{-1}$.}
    \label{fig:Bk_val_fid}
\end{figure}

\begin{figure}[h!]
    \centering
    \includegraphics[width=1.0\linewidth]{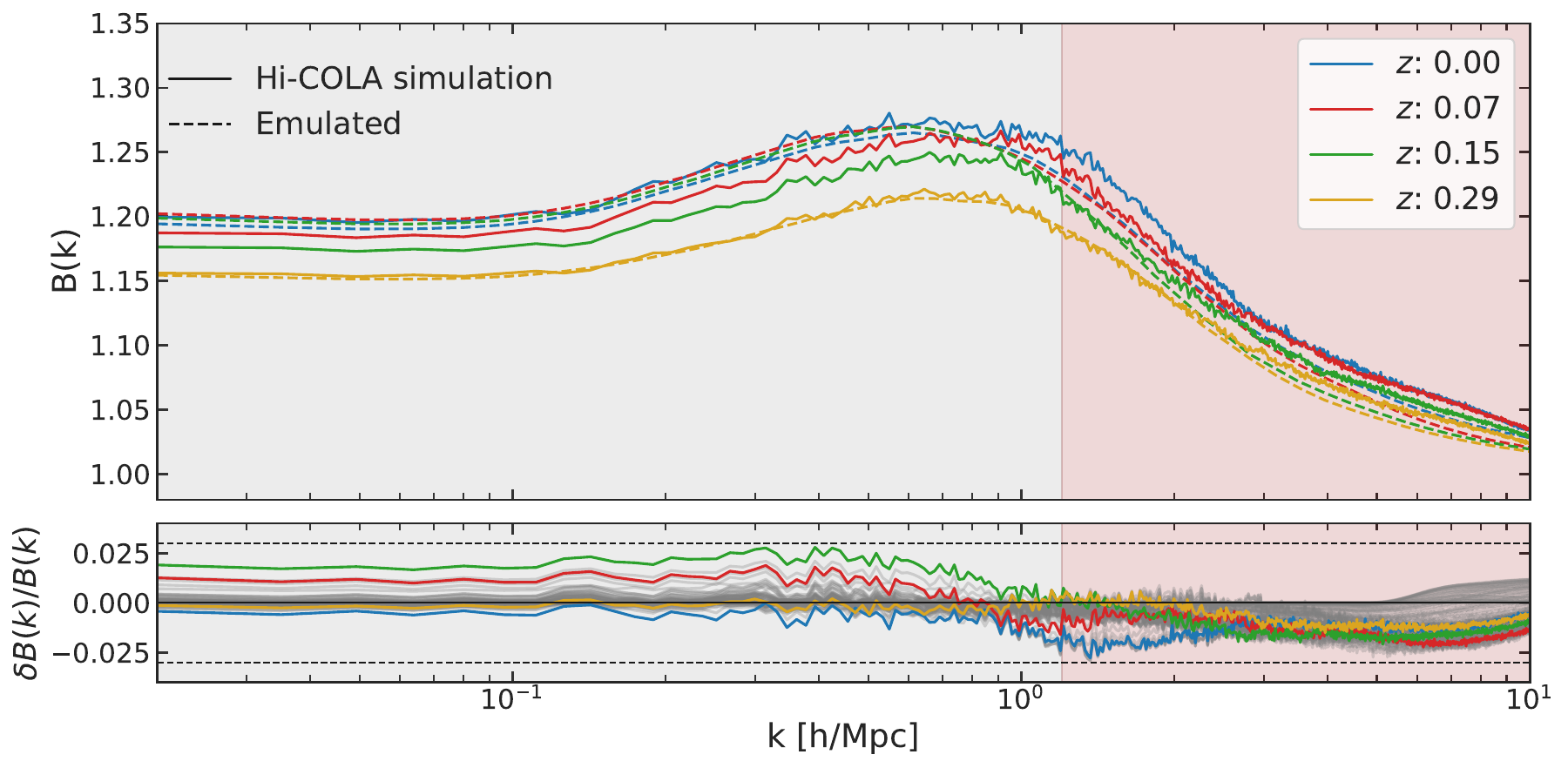}
    \caption{Top panel: Boosts predicted by the Cubic Galileon emulator (dashed lines) and the \HiCOLA{} boosts (solid lines) for the first validation set described in Section \ref{sec:Emu_results} with $f_{\phi}=1$ at a range of redshifts up to $z=0.29$. Bottom panel: relative error between the emulator prediction and the \HiCOLA{} boosts for the same validation set and redshift range. In gray, we show boost residuals at the remaining redshifts for which there was validation data (which includes redshifts up to $z=49$). The emulator is trained on a set smoothed with a Savitzky–Golay filter (window length 13, polynomial order 3), while the validation set was not smoothed.}
    \label{fig:Bk_val_fid_zdependence}
\end{figure}

Figure \ref{fig:Bk_val_fid} shows the comparison for the first validation set, in which only the parameter $f_\phi$ is varied while all other cosmological parameters are fixed to our fiducial cosmology. The agreement improves as $f_\phi$ decreases, i.e. as the GR limit is approached and the boosts become progressively flatter. This behaviour is common to emulators of this type \citep{fiorini2023fastproductioncosmologicalemulators, Ramachandra_2021}. The emulator predictions closely follow the \HiCOLA{} boosts at $z=0$. The corresponding relative errors are shown in the lower panel. We find that the emulator achieves $\sim 1\%$ accuracy for $k \lesssim 0.9\, h\,\text{Mpc}^{-1}$ (with less than 2\% error for 3$\sigma$ of the validation sample for $k \lesssim1.2 \,h\,\text{Mpc}^{-1}$), despite the validation data not being smoothed with the Savitzky-Golay filter (unlike for the training set). When looking at the emulator performance over the full range of redshifts (see Figure \ref{fig:Bk_val_fid_zdependence}), we additionally find that the emulator remains accurate at $< 3\%$ for $k \lesssim \,1.2 h\,\text{Mpc}^{-1}$. This demonstrates that the emulator generalizes well across variations in $f_\phi$ in regions of parameter space we expect to sample densely during parameter inference.

\begin{figure}[h!]
    \centering
    \includegraphics[width=1.0\linewidth]{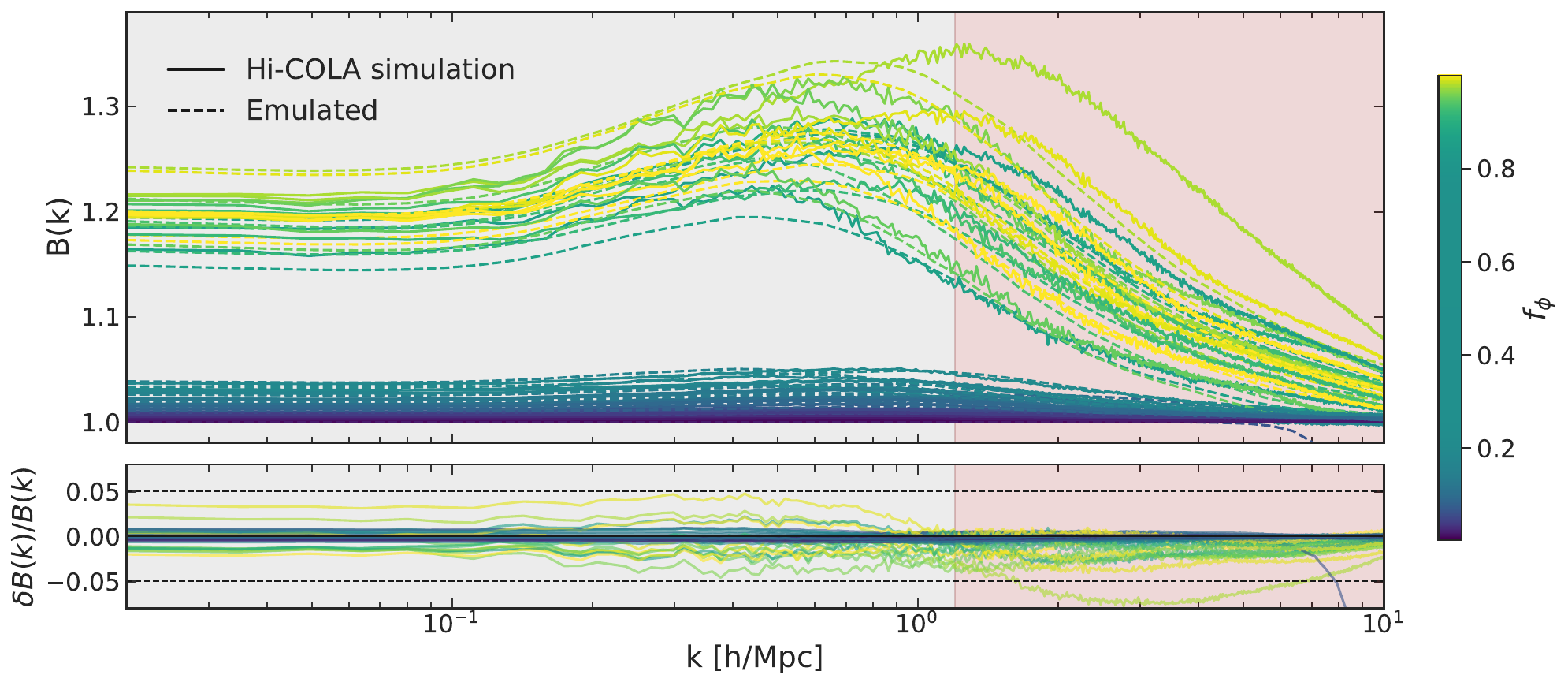}
    \caption{Top panel: Boosts predicted by the Cubic Galileon emulator (dashed lines) and the \HiCOLA{} boosts (solid lines) at $z=0$ for the validation set shown in Fig. \ref{fig:Validation_param_set}. The clustering into two groups (large vs. small boost) reflects the underlying validation values of $f_\phi$, which lie close to either 0 or 1 (see Fig. \ref{fig:Validation_param_set}). Bottom panel: relative error between the emulator prediction and the \HiCOLA{} boosts at $z=0$ for the same validation set. The emulator is trained on a set smoothed with a Savitzky–Golay filter (window length 13, polynomial order 3), while the validation set was not smoothed. 68\% of the emulated data has sub-percent errors, 95\% has an error $<2.5\%$, and 99.7\% at $<5\%$ for $k < 1.2 h\,\text{Mpc}^{-1}$.}
    \label{fig:Bk_val_edges}
\end{figure}

\begin{figure}[h!]
    \centering
    \includegraphics[width=1.0\linewidth]{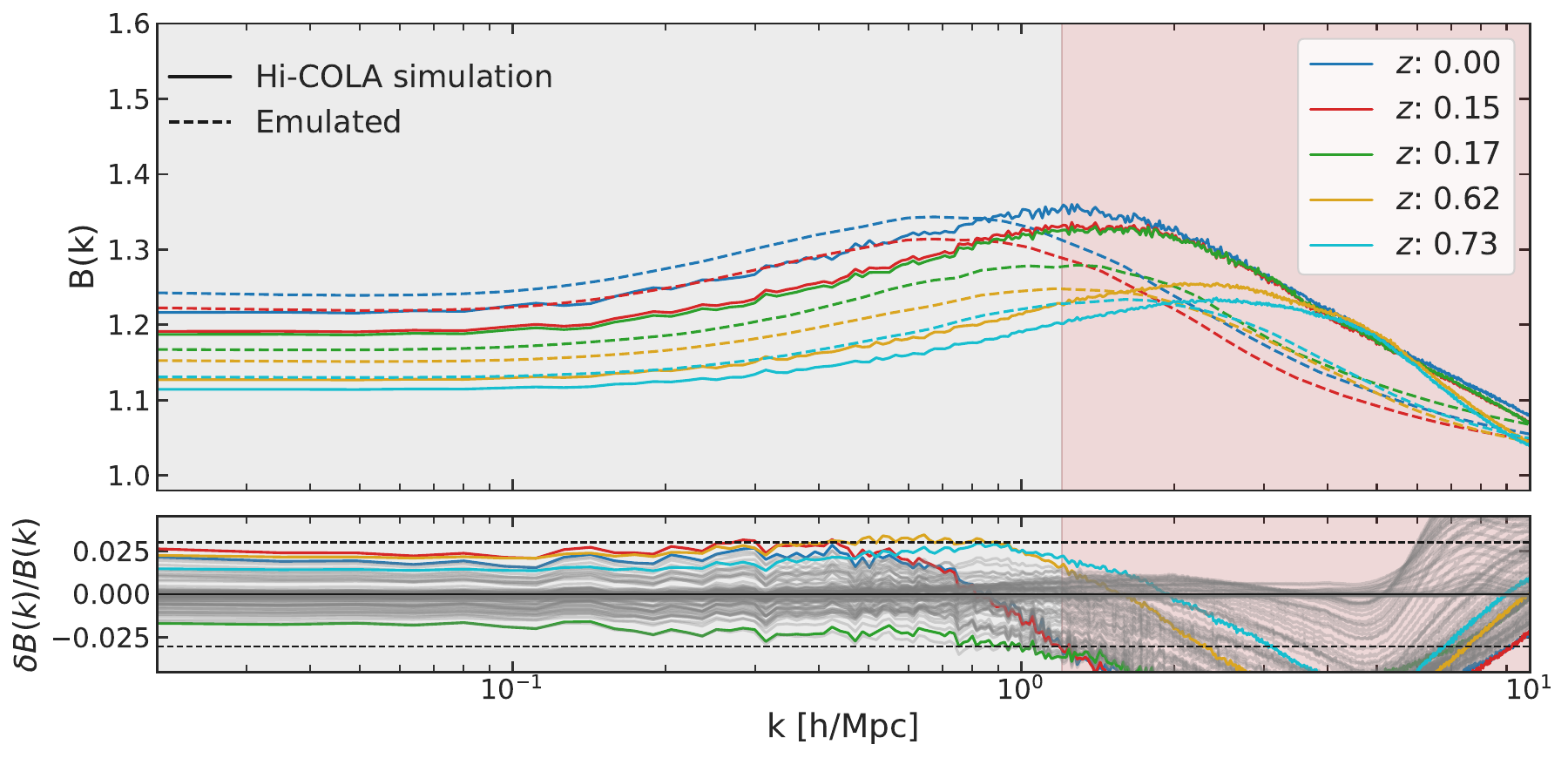}
    \caption{Top panel: Boosts predicted by the Cubic Galileon emulator (dashed lines) and the \HiCOLA{} boosts (solid lines) for a point in the second validation set with $\{\Omega_m = 0.284, \Omega_b=0.0600, h= 0.610, n_s=0.871, A_s = 1.479\times 10^{-09}, f_\phi=0.974\}$, for a range of redshifts up to $z=49$. Bottom panel: relative error between the emulator prediction and the \HiCOLA{} boosts for the same validation set and redshift range. In gray, we show boost residuals at the remaining redshifts for which there was validation data. The emulator is trained on a set smoothed with a Savitzky–Golay filter (window length 13, polynomial order 3), while the validation set was not smoothed.}
    \label{fig:Bk_val_edges_zdependence}
\end{figure}

Figure \ref{fig:Bk_val_edges} presents the results for the second validation set, which samples both the centre and edges of the parameter space. As expected, the errors increase toward the edges of the parameter volume, where the training data are sparser. Nevertheless, the performance remains good: 68\% of the predictions exhibit sub-percent accuracy ($<0.5\%$), 95\% of the predictions have errors below $2.5\%$, and 99.7\% below $5\%$ for $k < 1.2 h\,\text{Mpc}^{-1}$ at $z=0$. We also assess the emulator's performance across the full redshift range (see for example Figure \ref{fig:Bk_val_edges_zdependence}). Naively, one might expect the accuracy to improve monotonically with redshift, since the boost approaches unity as $z$ increases, and the modified gravity signal tends to the $\Lambda$CDM initial conditions. In practice, this improvement doesn't occur at intermediate redshifts: up to $z \sim 0.7$, the errors are comparable to those at $z=0$. This is partly attributable to the validation power spectra being computed with a larger number of timesteps than the emulator was trained on so that the expected improvement in accuracy is balanced out by the redshift interpolation error. We conclude that the emulator reproduces the boost well across redshift with a constant accuracy at low-to-intermediate z, beyond which the accuracy increases with redshift. Overall,  68\% of the predictions have an error $\lesssim 1.2\%$, 95\% below $2.5\%$, and 99.7\% below $\sim5\%$ for $k < 1.2 \,h\,\text{Mpc}^{-1}$. 

Overall, these results indicate that the emulator reliably reproduces \HiCOLA{} boosts across the relevant parameter space, with highest accuracy near the centre, and significant but sparse degradation toward the boundaries. This reduction in accuracy occurs primarily in sparsely sampled and physically unrealistic regions of parameter space, and is not expected to significantly impact cosmological parameter inference. In particular, we show that the emulator is accurate relative to \HiCOLA{} simulations to enable unbiased parameter estimation for LSST Y1 $3\times2$pt data with realistic scale cuts for physically motivated parameter values (see section \ref{sec:forecasts_results} below).

\subsection{Parameter inference for Stage IV-like data}\label{sec:forecasts}

We now forecast the constraints that LSST Y1 $3\times2$pt data could place on the Cubic Galileon model, using our emulator to model the modified gravity matter power spectrum, and propagating its predictions through to the observable two-point statistics. It takes between $2.5$ and $4$ hours on 76 cores to run an individual chain. We have described the methods and forecasting choices for this in Section \ref{sec:Inference_methods}. 

\subsubsection{Null test: forecasts for LSST Y1 relative to a $\Lambda$CDM universe}\label{sec:forecasts_null}

We expect a well-behaved pipeline should return an upper limit on the modified gravity parameter rather than a spurious detection when constraining a simulated universe with $f_\phi=0$ (i.e. a $\Lambda$CDM universe, to which we expect a realistic scenario of a future LSST Y1 3$\times$2pt analysis to be close). A preference for $f_\phi > 0$ in a GR universe would indicate model misspecification, for example from residual emulator inaccuracies being absorbed into the modified gravity sector. We therefore perform a null test, following the approach of recent Stage IV forecasts for $f(R)$ gravity \citep{Euclid2025}. We generate a noiseless synthetic $3\times2$pt datavector in which the matter power spectrum is computed for GR using \texttt{CCL}. We then analyse it with the Cubic Galileon emulator, using the same sampler settings, baryonic scale cuts and maximum wavevector for galaxy clustering, $k_{\text{max}} = 0.2\,h\,\text{Mpc}^{-1}$, as in our baseline analysis (Section \ref{sec:Inference_methods}). We also impose {\it Planck} priors on $\omega_b$ and $n_s$.

\begin{figure}
    \centering
    \includegraphics[width=0.8\linewidth]{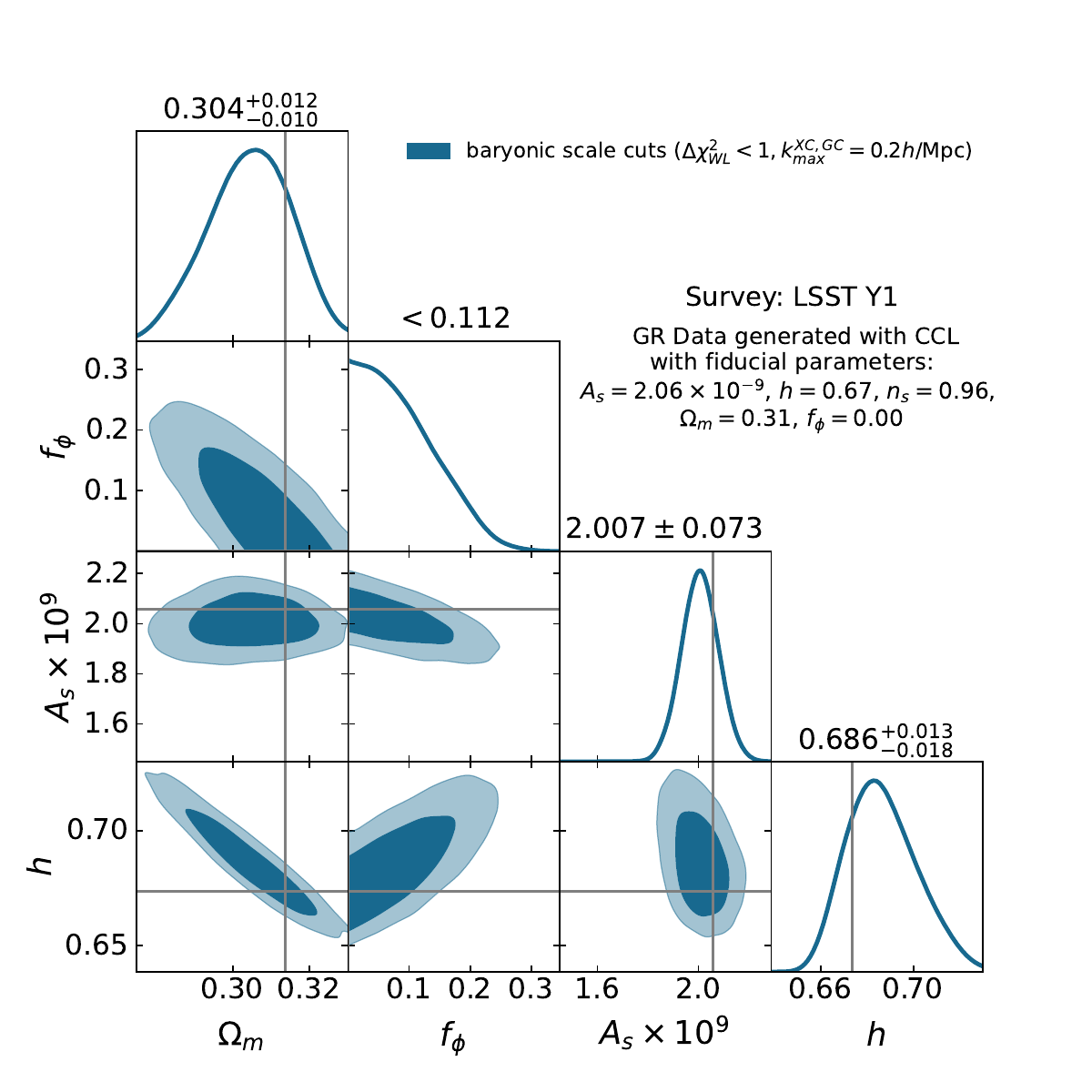}
    \caption{Posterior constraints from an LSST Y1 $3\times2$pt forecast using mock data generated in GR, analysed with the Cubic Galileon emulator and {\it Planck} priors. The recovered cosmological and nuisance parameters are unbiased, and the posterior on $f_\phi$ is consistent with zero, giving $f_\phi < 0.202$ at 95\% CL.}
    \label{fig:3x2pt_corner_GR}
\end{figure}

Figure \ref{fig:3x2pt_corner_GR} shows the resulting posteriors. The recovered cosmological and nuisance parameters are consistent with their input values, and we find no preference for a departure from GR. The marginalised posterior on $f_\phi$ peaks at the $\Lambda$CDM boundary, $f_\phi = 0$, and we obtain upper limits of
\begin{equation}
    f_\phi < 0.112 \;\; (68\%\,\text{CL}), \qquad f_\phi < 0.202 \;\; (95\%\,\text{CL}).
\end{equation}
Our pipeline therefore passes the null test. We note here that these bounds depend on our choice of prior on $f_\phi$, the {\it Planck} prior on $n_s$ and $\omega_b$, and our systematic modelling choices, and because of the latter two they are optimistic for our particular scale cut choice.

\subsubsection{Forecasts for LSST Y1 relative to a \HiCOLA{} simulated universe}\label{sec:forecasts_results}
 
\begin{figure}
    \centering
    \includegraphics[width=0.8\linewidth]{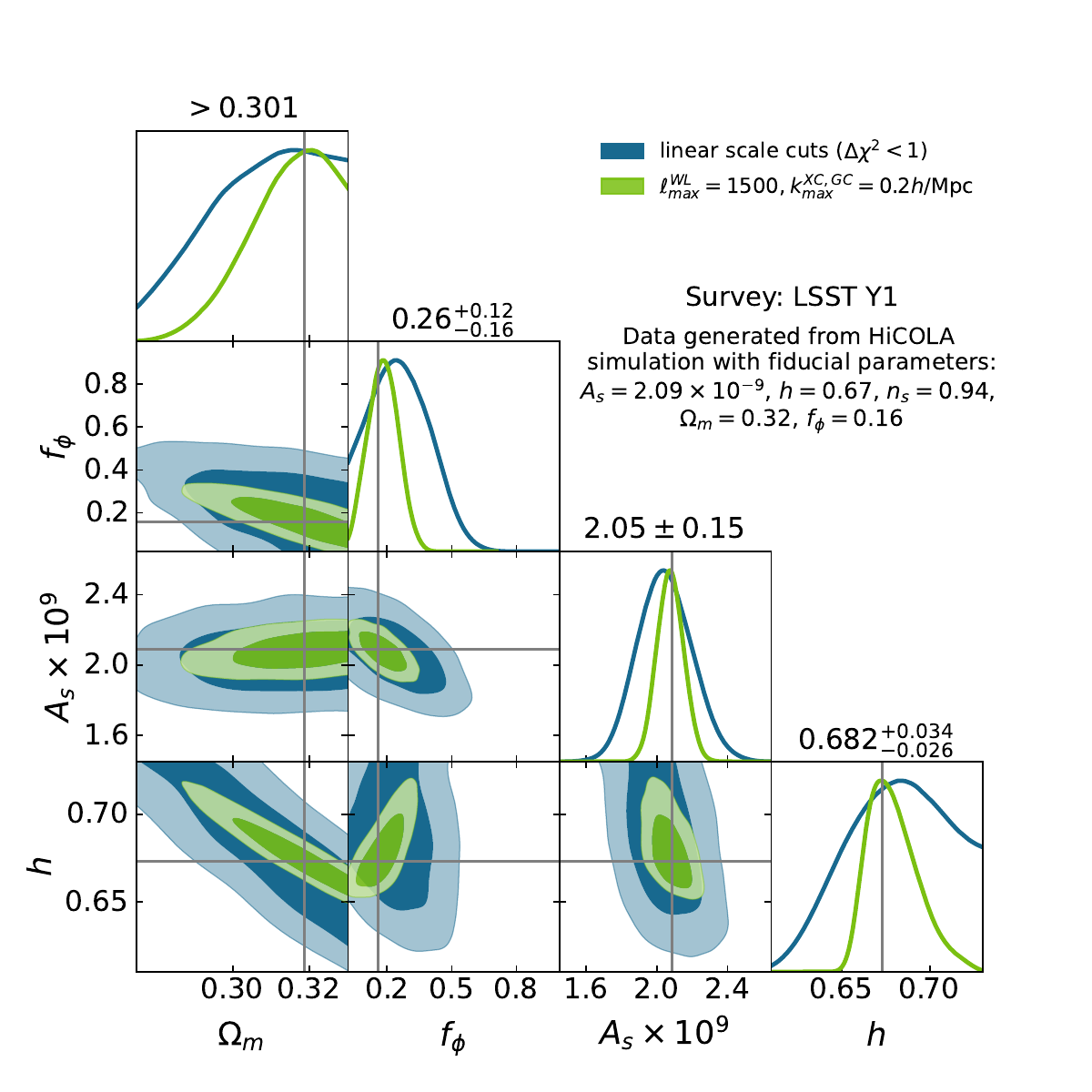}
    \caption{Posterior constraints from an LSST Y1 3$\times$2pt forecast using mock data generated from \HiCOLA{} for a validation point located near the fiducial cosmology within the LHS (with $f_\phi \approx 0.16$). The theoretical model uses the Cubic Galileon emulator. No significant bias is observed in either cosmological or nuisance parameters.}
    \label{fig:3x2pt_corner_i=45}
\end{figure}

We sample our posterior parameter space using the \texttt{nautilus} \citep{nautilus} sampler, with minimum effective sample size of $20000$ and a maximum live set evidence fraction of 0.01. Figure \ref{fig:3x2pt_corner_i=45} shows the posteriors for our analysis with a simulated noiseless datavector which uses a matter power spectrum generated with \HiCOLA, for a validation point with $f_\phi \approx 0.16$ (a more realistic and as of yet unconstrained value of $f_\phi$), with a maximum wavevector for galaxy clustering angular power spectra of $k_{\text{max}} = 0.2 h\,\text{Mpc}^{-1}$. The emulator produces unbiased constraints on the cosmological and modified gravity parameters for both the baryonic cuts described in Section \ref{sec:Inference_methods} and cuts at $\ell_{\rm max}=1500$, and increases the significance of the detection in $f_\phi$ relative to linear-only modelling from $1.9\sigma$ to $2.6 \sigma$. 
 
\begin{figure}
    \centering
    \includegraphics[width=0.8\linewidth]{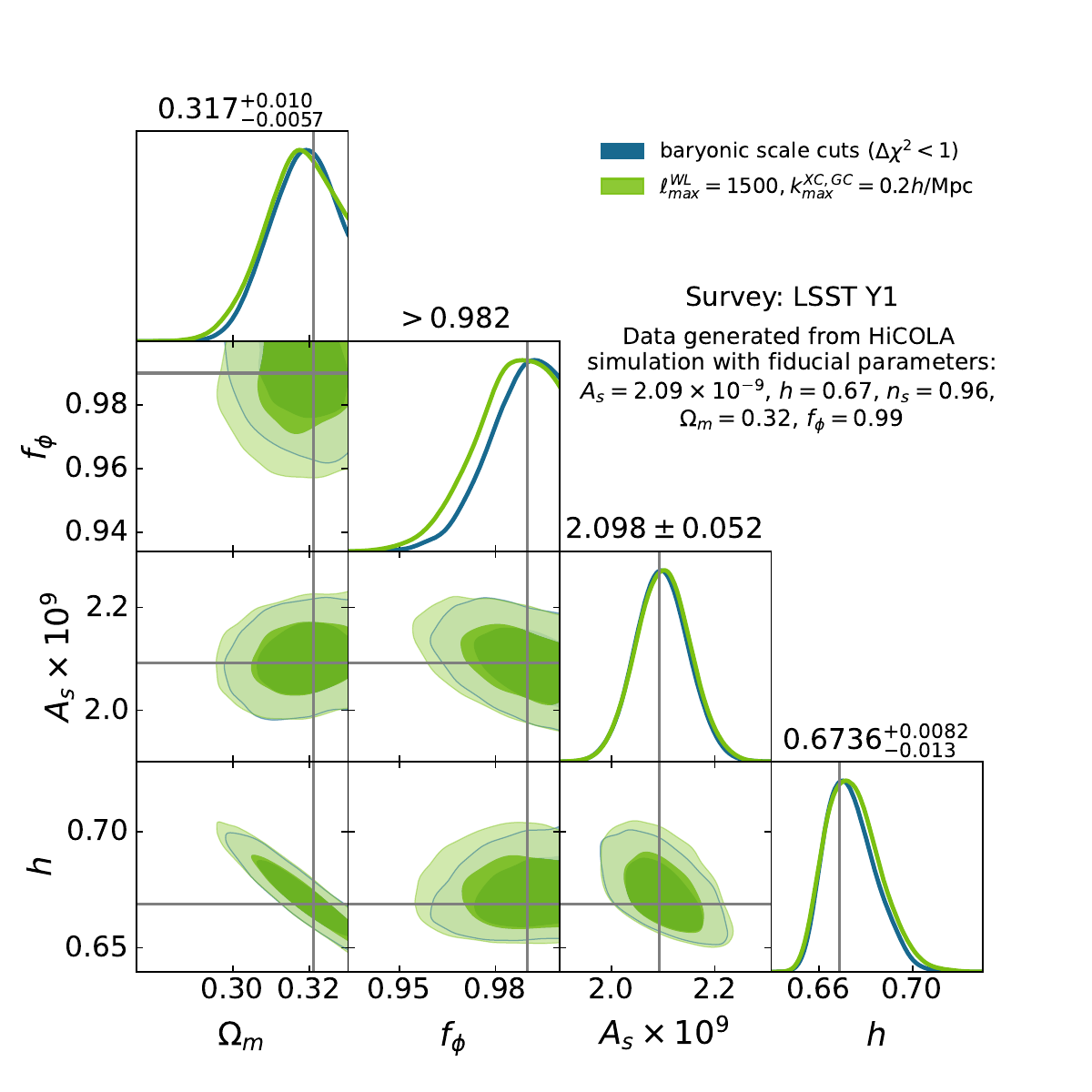}
    \caption{Posterior constraints from an LSST Y1 3$\times$2pt forecast using mock data generated from \HiCOLA{} for a validation point located near the fiducial cosmology within the LHS (with $f_\phi = 0.99$). The theoretical model uses the Cubic Galileon emulator. No significant bias is observed in either cosmological or nuisance parameters.}
    \label{fig:3x2pt_corner_i=24}
\end{figure}

We also test the robustness of the emulator under less conservative scale cuts for a stronger modification to $\Lambda$CDM. Because we are assessing the bias introduced by the emulator, we design a worst-case scenario by changing two things at once relative to the previous test. We increase $f_\phi$, increasing the MG signature and the error on the boost, and we relax the scale cuts, introducing scales where the emulator inaccuracies are largest. Figure \ref{fig:3x2pt_corner_i=24} shows the posteriors for our analysis with a simulated noiseless datavector which uses a \HiCOLA{} matter power spectrum with $f_\phi =0.99$, with a maximum wavevector for galaxy clustering angular power spectra of $k_{\text{max}} = 0.2 h\,\text{Mpc}^{-1}$. Both with baryonic cuts for shear, and with uniform scale cuts for shear of $\ell_{\text{max}}^{\kappa \kappa} = 1500$, we recover unbiased constraints in our cosmological and nuisance parameters.  The emulator remains reliable for a large departure from $\Lambda$CDM and under more aggressive scale cuts than our baseline analysis. We caution that this conclusion depends on our scale cut scheme. A nulling scheme such as the BNT transform \cite{Taylor_2021}, which retains information from smaller angular scales, could expose inaccuracies in the forecast masked by the decreased statistical significance of the detection after conservative scale cuts. In that case, the unbiased recovery reported here would not necessarily hold.

\subsubsection{Forecasts for LSST Y1 relative to full N-body simulations}\label{sec:forecasts_results_nbody}

The forecast in Section \ref{sec:forecasts_results} isolates the emulator construction error and demonstrates that the emulator recovers unbiased parameters when the mock data are generated with \HiCOLA. It characterises the error in the emulator relative to the \HiCOLA{} simulations it was trained on, but it doesn't capture the error inherent in the COLA approximation, which trades small-scale accuracy for speed  (\cite{Wright_2023} finds that \HiCOLA{} simulations are valid at the 2.5\% level down to $k_{\text{max}} = 1.2 h\,\text{Mpc}^{-1}$ at $z = 0$). To quantify this contribution to the total error budget of the Cubic Galileon emulator, we compare \HiCOLA{} with a high-resolution N-body simulation run with \texttt{EFT-RAMSES} \cite{Woodcock:2026}, an adaptive mesh refinement code for modified gravity built on \texttt{RAMSES}. To do this, we use the power spectra for a Cubic Galileon model on the tracker solution with $f_\phi = 1$ and $\Omega_m=0.3138, \Omega_b=0.0493, h=0.6736, n_s=0.9649,\sigma_8=0.8143$ for two phase-reversed simulations, together with matched GR runs sharing the same initial conditions, so that the simulated boost is computed consistently with the emulator. While Cubic Galileon without a cosmological constant has been ruled out experimentally, $f_\phi = 1$ is the maximal departure from $\Lambda$CDM in our parameter range, ensuring we consider the most stringent test of the COLA approximation. Here, we apply a more demanding test than that in Section \ref{sec:forecasts_results}, by analysing a mock datavector built from the full N-body \texttt{EFT-RAMSES} simulation, so that the COLA approximation error now enters as a potential source of bias on the inferred parameters.

\begin{figure}
    \centering
    \includegraphics[width=1.0\linewidth]{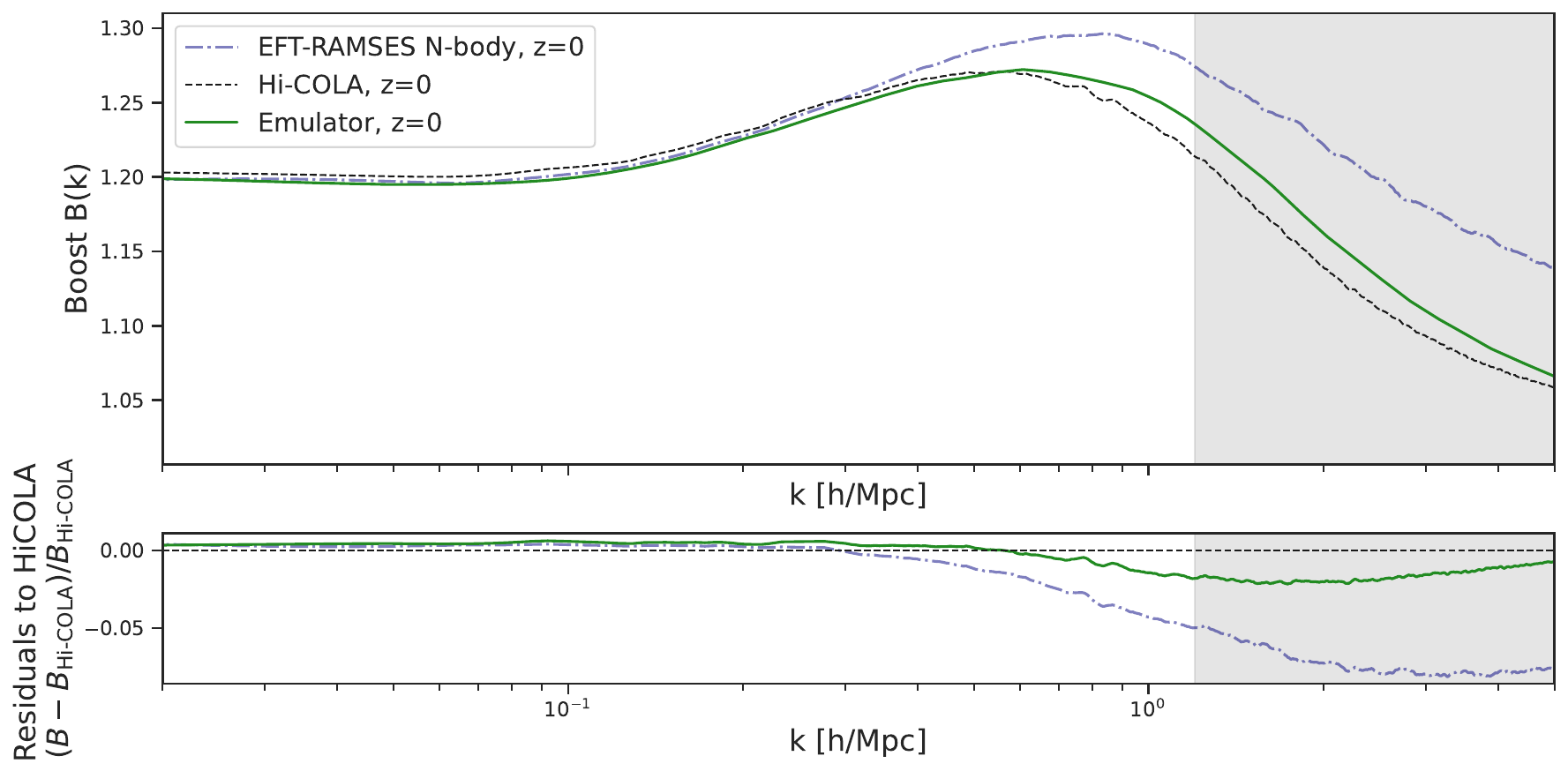}
    \caption{Top panel: the \texttt{EFT-RAMSES} boost, the \HiCOLA{} boost, and the emulator prediction at $z=0$ for $f_\phi=1$. Bottom panel: relative error of \HiCOLA{} vs \texttt{EFT-RAMSES}, and \HiCOLA{} vs the emulator prediction.}
    \label{fig:Boost_comparison_Nbody}
\end{figure}

We construct the simulated datavector from the \texttt{EFT-RAMSES} Cubic Galileon power spectrum. The boosts and residuals for the \HiCOLA{} simulation relative to the \texttt{EFT-RAMSES} simulation and the emulator are displayed in Figure \ref{fig:Boost_comparison_Nbody}. 
First, we estimate the theoretical Cubic Galileon boost from the average of two realizations with phase-shifted ICs, $B(k,z) = \big(P^{CG}_{1} + P^{CG}_{2}\big)\big/\big(P^{\Lambda\text{CDM}}_{1} + P^{\Lambda\text{CDM}}_{2}\big)$, and smooth it with the same Savitzky–Golay filter (window length 13, polynomial order 3) applied to the training set. Outside the simulated range, we follow a similar extrapolation procedure as for the emulator: for $z > 2$ we replace the boost with its linear theory prediction until $z=12$, and then set the boost to 1. For $k\geq5 h\,\text{Mpc}^{-1}$ we use the power-law form $B(k) = A k^{-p} + 1$ with $p \geq 0$, ensuring the boost decays smoothly towards unity on small scales. The boost is then multiplied by a $\Lambda$CDM nonlinear matter power spectrum estimated with halofit (to ensure consistency with our model vector) to build the 3$\times$2pt simulated datavector through Equation \eqref{eq:C_ell_general}. We then estimate the datavector as outlined in Section \ref{sec:Inference_methods}.

\begin{figure}
    \centering
    \includegraphics[width=0.8\linewidth]{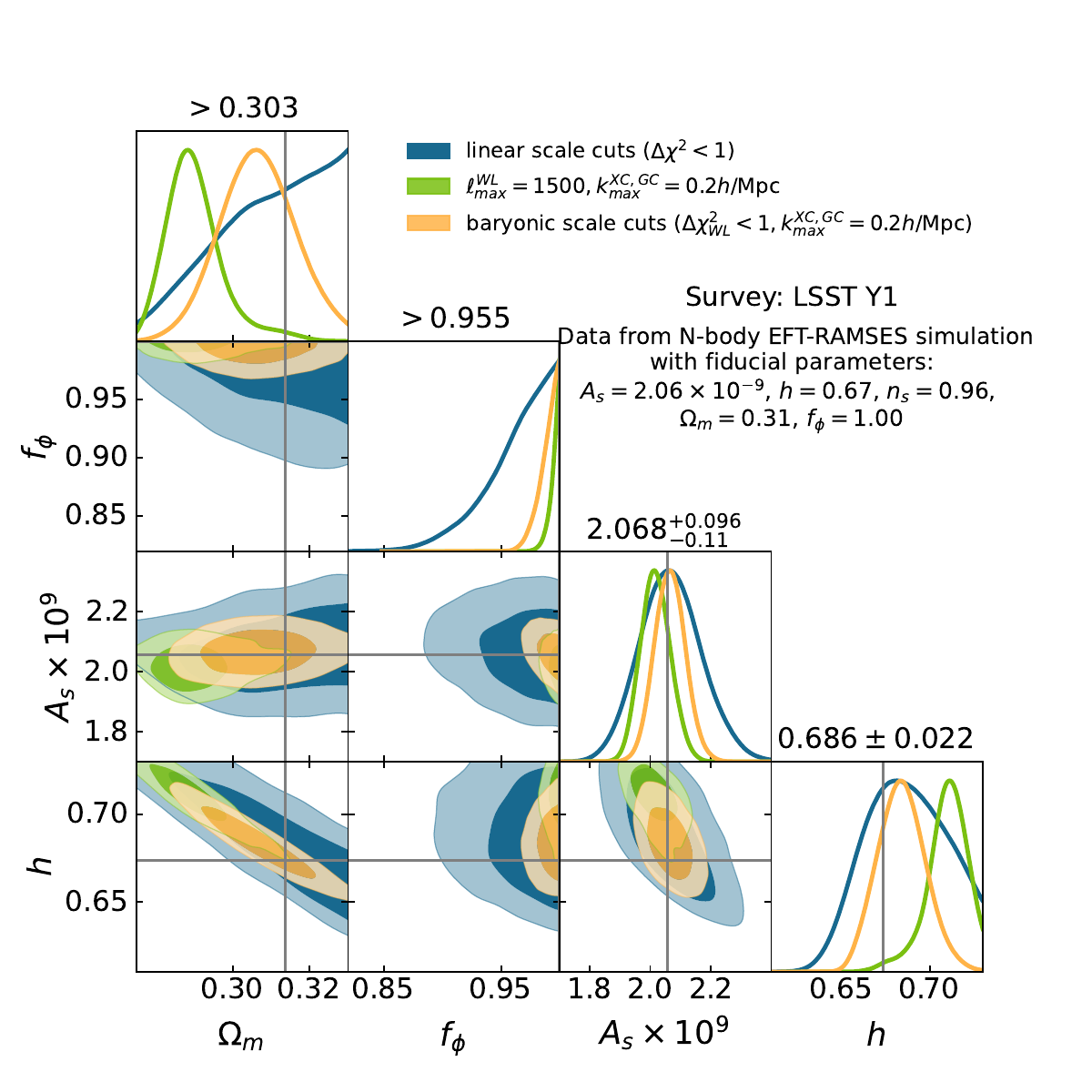}
    \caption{Posterior constraints from an LSST Y1 3$\times$2pt forecast using mock data generated from an \texttt{EFT-RAMSES} N-body simulation near the fiducial cosmology (with $f_\phi = 1$). The theoretical model uses the Cubic Galileon emulator. No significant bias is observed when scale cuts are baryonic (as defined in section \ref{sec:Inference_methods}), but significant ($>2\sigma$) bias is introduced for less conservative scale cuts methods ($\ell_{\text{max}}^{\kappa \kappa} = 1500$).}
    \label{fig:3x2pt_corner_Nbody}
\end{figure}

Using the \texttt{EFT-RAMSES} $3\times2$pt vector as our simulated data, we carry out the same \texttt{nautilus}-sampled 3$\times$2pt inference as in Sections \ref{sec:forecasts_null} and \ref{sec:forecasts_results}, using the Cubic Galileon emulator as the theoretical model. Figure \ref{fig:3x2pt_corner_Nbody} shows the posterior distribution for the parameters in this analysis. We can see that the cosmological parameters are significantly biased when using $\ell_{\rm max}^{\rm WL} = 1500$. However, we note that $f_\phi$ itself remains consistent with its fiducial value in this case. This is likely an artefact of the prior boundary: the fiducial $f_\phi = 1$ sits at the upper edge of the emulator's training range, so any bias towards larger $f_\phi$ is truncated by the prior rather than being absorbed into that parameter.

We now want to understand if the bias is due to a breakdown of the COLA approximation, or a loss of emulator accuracy at small scales. The total emulator error relative to full N-body propagated to parameter constraints comes from two contributions: the emulator-vs-\HiCOLA{} error (which affects the results described in Section \ref{sec:forecasts_results}), and the \HiCOLA-vs-N-body (i.e. COLA approximation) error. 
We have shown in Section \ref{sec:forecasts_results} that when the mock data are generated from \HiCOLA{} at $f_\phi = 0.99$ and analysed at a fixed clustering cut of $\ell_{\text{max}} = 1500$, the emulator recovers the input cosmology and $f_\phi$ without significant bias for physical values of the cosmological parameters (Figure \ref{fig:3x2pt_corner_i=24}, in green). When the mock data are instead generated from the \texttt{EFT-RAMSES} N-body simulation and analysed at the same fixed $\ell_{\text{max}} = 1500$, the recovered cosmology is biased at $>2\sigma$ (Figure \ref{fig:3x2pt_corner_Nbody}, in green). Taken together, these results show that the emulator cannot reproduce the behaviour of the N-body simulation at these scales because it inherits the \HiCOLA{} small-scale behaviour. These results therefore show that the dominant limitation is not the emulator interpolation but the COLA approximation on small scales. 

When the same \texttt{EFT-RAMSES} mock data are analysed with the more conservative baryonic scale cuts (eq. \ref{eq:clustering_cut}, with $k_{\text{max}} = 0.2\ h\,\text{Mpc}^{-1}$), the emulator recovers an unbiased cosmology during parameter inference (Figure \ref{fig:3x2pt_corner_Nbody}, in orange). The scales at which the COLA approximation breaks down lie beyond those retained after baryonic cuts, so the emulator is fit for purpose for LSST Y1 under a realistic but conservative analysis. For a less conservative analysis, we advise to test the emulator against N-body simulations beforehand for a more realistic setup (including systematics like intrinsic alignment modelling and a more realistic photo-$z$ uncertainty), which we expect might be more forgiving than this low-parameter simplified forecast. 

Overall, we show that the baryonic scale cuts required to control for the systematic of baryonic feedback also suppress the COLA-induced bias. Using less conservative scale cuts (keeping scales up to $\ell = 1500$) exposes the COLA small-scale approximation error. The emulator is therefore sufficient for unbiased Cubic Galileon parameter inference with LSST Y1 3$\times$2pt data if conservative scale cuts are applied. 

\section{Conclusions}\label{sec:Conclusions}

We created an emulator over 5 parameters, $\Omega_{m0}, n_s, A_s, h$ and modified gravity parameter $f_\phi$, for the boost of a Cubic Galileon model whose background follows the tracker solution, trained on \HiCOLA{} simulations. We validated this emulator against both \HiCOLA{} and full N-body (\texttt{EFT-RAMSES}) simulations, comparing the predicted boosts directly as well as forecasting constraints for a simulated LSST Y1 $3\times2$pt analysis. The emulator and forecasting code are publicly available\footnote{\url{https://github.com/LSSTDESC/CubicGalileonEmu}}$^{,}$\footnote{\url{https://github.com/CarolaZano/CubicGalileon_Forecasts}}.

The emulator reliably reproduces \HiCOLA{} boosts, with highest accuracy near the centre of the parameter hypervolume ($\sim 1\%$ accuracy for $k \lesssim 0.9 h\,\text{Mpc}^{-1}$), and larger occasional degradation toward the boundaries, where the training data are sparser (with most errors remaining below $5\%$ for $k < 1.2 h\,\text{Mpc}^{-1}$). The emulator is accurate enough relative to \HiCOLA{} simulations to enable unbiased parameter estimation for LSST Y1 $3\times2$pt data, for physically motivated parameter values, even under optimistic scale cuts ($\ell^{\kappa \kappa} \leq 1500$ and $k^{\rm GC,XC} \leq 0.2 h\,\text{Mpc}^{-1}$). 

We also find that the boost deviates more substantially when compared against a high-resolution N-body (\texttt{EFT-RAMSES}) simulation, which we treat as the benchmark in our analysis. The \HiCOLA{} and emulator boosts differ from the \texttt{EFT-RAMSES} boost by $\sim 5\%$ at $k \approx 1.2\ h\,\text{Mpc}^{-1}$. This shows that the COLA approximation rather than the emulator interpolation is the dominant source of error on small scales. When \texttt{EFT-RAMSES} mock data are analysed with  the optimistic scale cuts above, the inferred cosmological parameters are biased at $\geq 2\sigma$. When more conservative baryonic scale cuts are applied (eq. \ref{eq:clustering_cut}, with $k_{\rm max} = 0.2 h\,\text{Mpc}^{-1}$), the emulator recovers an unbiased cosmology during parameter inference for 3$\times2$pt LSST Y1-like data. The emulator can therefore produce unbiased posteriors for our Cubic Galileon model with LSST Y1-like 3$\times$2pt data under conservative baryonic cuts. We stress that the accuracy demonstrated here applies to the specific LSST Y1 3$\times$2pt setup and scale cuts we adopted. It should not be read as validation for Stage IV analyses in general.

More specifically, this conclusion depends on the scale cut scheme: a nulling approach such as the BNT transform \cite{Taylor_2021}, which retains information from smaller angular scales, could expose inaccuracies masked by our conservative cuts.  We also note that, since baryonic feedback in MG remains poorly understood, we used DMO power spectra modelling, with scale cuts which impose a strict removal of $\ell$ bins affected by baryonic physics, rather than introducing additional parameters for baryonic physics modelling. Finally, our forecast is a simplified, low-dimensional setup that neglects systematics such as intrinsic alignments and adopts a simplified photo-$z$ treatment. We expect this last simplification to be conservative as far as the emulator is concerned: a more complete analysis with additional nuisance parameters would widen the posteriors and introduce degeneracies that could further mitigate the modelling biases. We recommend that these forecasts be repeated for the specific analysis setup of interest, with the full parameter set and the chosen scale cuts scheme, to confirm the emulator's accuracy under those specifications.

This Cubic Galileon model is a good target for emulator development with \HiCOLA, as it provides an alternative theoretically motivated model to Hu-Sawicki $f(R)$ and nDGP, and it modifies the cosmological background. It is, however, of limited interest as an alternative to dynamical dark energy: it has been ruled out observationally at $f_\phi = 1$, and it does not cross the phantom divide.  Future work will therefore aim to extend the emulated parameter space to Horndeski theories that cross the phantom divide, addressing more directly the MG modelling required to address the dynamical dark energy signals now emerging from the data.  Additionally, in future emulation with \HiCOLA, we will improve on the current techniques by using a larger training set spanning wider parameter ranges (for example in $\Omega_{m0}$), additionally provide LSST Y10 forecasts, and aim to introduce less conservative scale cuts for baryonic physics modelling (e.g. \cite{Huang_2019} or \cite{Taylor_2021}), potentially introducing propagation of the \texttt{EFT-RAMSES}-to-\HiCOLA{} error into the covariance matrix. 

More broadly, this work demonstrates that the combination of \HiCOLA{} and emulation is a viable and fast pipeline for under-explored Horndeski theories towards developing nonlinear modelling tools useful for Stage IV analyses, providing a route to systematically test a much wider class of gravity models than the simulation infrastructure has so far allowed.

\section*{Acknowledgements}

This work has undergone internal review by the LSST Dark Energy Science Collaboration; the internal reviewers were Sankarshana Srinivasan and Georgios Valogiannis. CMAZ is supported by the Philip Robinson Cosmology PhD studentship. ASG is supported by the European Union (ERC StG, LSS BeyondAverage, 101075919). TB is supported by ERC Starting Grant SHADE (grant no.\,StG 949572) and by a Royal Society University Research Fellowship (grant no.URF\textbackslash R\textbackslash 231006). CDL is supported by the Science and Technology Facilities Council (STFC) [grant No. UKRI1172]. We acknowledge travel support provided by STFC for UK participation in the Rubin LSST, through grant ST/X001334/1.

Work at Argonne National Laboratory was supported by the U.S. Department of Energy, Office of High Energy Physics. Argonne, a U.S. Department of Energy Office of Science Laboratory, is operated by UChicago Argonne LLC under contract no. DE-AC02-06CH11357. Part of this work was done on Improv, a CPU system located at the Laboratory Computing Resource Center (LCRC) of Argonne National Laboratory.

The DESC acknowledges ongoing support from the Institut National de 
Physique Nucl\'eaire et de Physique des Particules in France; the 
Science \& Technology Facilities Council in the United Kingdom; and the
Department of Energy and the LSST Discovery Alliance
in the United States.  DESC uses resources of the IN2P3 
Computing Center (CC-IN2P3--Lyon/Villeurbanne - France) funded by the 
Centre National de la Recherche Scientifique; the National Energy 
Research Scientific Computing Center, a DOE Office of Science User 
Facility supported by the Office of Science of the U.S.\ Department of
Energy under Contract No.\ DE-AC02-05CH11231; STFC DiRAC HPC Facilities, 
funded by UK BEIS National E-infrastructure capital grants; and the UK 
particle physics grid, supported by the GridPP Collaboration.  This 
work was performed in part under DOE Contract DE-AC02-76SF00515.

For the purpose of open access, the author has applied a Creative Commons Attribution (CC BY) licence to any Author Accepted Manuscript version arising from this submission.

\paragraph{Author Contributions.}

CMAZ ran emulator training and validation simulation suites, performed the analysis tied to Sections \ref{sec:DataGeneration}, \ref{sec:Inference_methods}, and \ref{sec:Emu_results} (including writing and running the forecasting code) and wrote a significant portion of the text of this paper.
ASG worked on the modelling of the on-tracker Cubic Galileons, assisted with the model selection for creating the emulator training and validation sets, and assisted with the running of \HiCOLA{}. ASG also assisted with the measurements of 2-point matter power spectra, and worked on the analysis and comparison of results with the \texttt{EFT-RAMSES} code. ASG contributed to the writing and revising of the manuscript.
NR developed the emulator, played a major role in designing the simulation suite and the forecasting analysis, wrote Section \ref{sec:Emulator_development}, and contributed to writing and revising the rest of the manuscript.
TB worked on the modelling of the on-tracker Cubic Galileons, advised on running \HiCOLA{} and on the simulation specifications, provided scientific and technical guidance particularly on the work in Sections \ref{sec:Theory_and_Sims}, \ref{sec:DataGeneration}, and \ref{sec:Emu_results}, and contributed to writing and revising the manuscript.
BF advised on running \HiCOLA{} and on the simulation specifications, and provided guidance on the emulator design, on the COLA code, on the modeling of the on-tracker Cubic Galileons, and on the comparison with N-body simulations.
DJ contributed to the early development of the emulation infrastructure and to the project's inception.
KH helped conceive and design the project, assisted with planning the preliminary simulation suite and developing the initial version of the emulator, and contributed to revising the manuscript.
CDL provided scientific and technical guidance particularly with regards to the work of Sections \ref{sec:DataGeneration}, \ref{sec:Inference_methods}, and \ref{sec:Emu_results}, and contributed to writing and revising the paper text.
BL provided access to the \texttt{EFT-RAMSES} simulation, gave scientific and technical guidance, and contributed to writing and revising the manuscript.
NŠ developed tools for redshift distributions and tomographic binning; DESC Builder.

\bibliographystyle{unsrt}
\bibliography{References}

\end{document}